\documentclass[11pt,twoside]{article}

\usepackage{a4}
\usepackage{amsmath}
\usepackage{amssymb}
\usepackage{amsthm}
\usepackage{amsfonts}
\usepackage{amstext}
\usepackage{amsopn}
\usepackage{graphicx}
\usepackage[latin1]{inputenc}
\usepackage{eufrak}
\newcommand\R{{\ensuremath {\mathbb R} }}
\newcommand\C{{\ensuremath {\mathbb C} }}
\newcommand\N{{\ensuremath {\mathbb N} }}
\newcommand\Z{{\ensuremath {\mathbb Z} }}

\renewcommand\S{{\ensuremath {\EuFrak{S}} }}
\renewcommand\phi{\varphi}

\newcommand{\bx}{\mathbf{x}}
\newcommand{\by}{\mathbf{y}}
\newcommand{\Hh}{\mathbb{H}}
\newcommand{\alp}{\boldsymbol{\alpha}}

\newcommand\ii{{\ensuremath {\infty}}}
\newcommand\pscal[1]{{\ensuremath{\langle #1 \rangle}}}
\newcommand{\norm}[1]{ \left| \! \left| #1 \right| \! \right| }

\newcommand\Tr[1]{{\ensuremath \mathrm{Tr}_{\C^4}\left(#1\right)}}
\newcommand{\normQ}[1]{\ensuremath{ | \! | #1 | \! |_{\mathcal{Q}}}}
\newcommand{\normR}[1]{\ensuremath{ | \! | #1 | \! |_{\mathcal{R}}}}
\newcommand{\normr}[1]{\ensuremath{ | \! | #1 | \! |_{\EuFrak{C}}}}
\newcommand{\normp}[1]{\ensuremath{ | \! | #1 | \! |_{\mathcal{Y}}}}
\newcommand{\normd}[1]{\ensuremath{ | \! | #1 | \! |_{\EuFrak{C}'}}}
\def\tr{\mathop{\rm tr}\nolimits} 
\def\str{\mathop{\rm tr}\nolimits} 

\newtheorem{thm}{Theorem}
\newtheorem{lemma}{Lemma}

\newtheorem{prop}[lemma]{Proposition}
\newtheorem{definition}{Definition}
\newtheorem{remark}{Remark}

\title{\bf Existence of a stable polarized vacuum in the Bogoliubov-Dirac-Fock approximation}
\author{\bf Christian HAINZL\footnote{Second address: Laboratoire de Math\'ematiques Paris-Sud. Bât. 425, 91 405 Orsay Cedex. France}, Mathieu LEWIN \& \'Eric S\'ER\'E}
\date{ }
\begin{document}
\maketitle

\vspace{-0.7cm}
\begin{center}
CEREMADE, UMR CNRS 7534, Université Paris IX Dauphine, Place du 
Maréchal De Lattre De Tassigny, 75 775 Paris, Cedex 16, France.

\bigskip

E-mail: {\it hainzl, lewin, sere@ceremade.dauphine.fr}
\end{center}

\medskip

\begin{abstract}
According to Dirac's ideas, the vacuum consists of infinitely many 
virtual electrons which completely fill up the negative part of the 
spectrum of the free Dirac operator $D^0$. In the presence of an 
external field, these virtual particles react and the vacuum becomes 
polarized.

In this paper, following Chaix and Iracane ({\it J. Phys. B}, 22, 
3791--3814, 1989), we consider the Bogoliubov-Dirac-Fock model, which 
is derived from no-photon QED. The corresponding BDF-energy takes the 
polarization of the vacuum into account and is bounded from below. A 
BDF-stable vacuum is defined to be a minimizer of this energy. If it 
exists, such a minimizer is solution of a self-consistent equation.

We show the existence of a unique minimizer of the BDF-energy in the 
presence of an external electrostatic field, by means of a 
fixed-point approach. This minimizer is interpreted as the polarized 
vacuum.
\end{abstract}

\section{Introduction}
The relativistic quantum theory of electrons and positrons is based 
on the free Dirac operator, which is defined by
\begin{equation}
D^0=-i\sum_{k=1}^3\alpha_k\partial_k+\beta :=-i\alp\cdot \nabla+\beta
\label{dirac_free}
\end{equation}
where $\alp=(\alpha_1,\alpha_2,\alpha_3)$ and
$$\beta=\left(\begin{matrix}
I_2 & 0\\ 0 & -I_2\\
\end{matrix}\right),\qquad
\alpha_k=\left(\begin{matrix}
0 & \sigma_k\\ \sigma_k & 0\\
\end{matrix}\right),$$
with
$$\sigma_1=\left(\begin{matrix}
0 & 1\\ 1 & 0\\
\end{matrix}\right), \qquad
\sigma_2=\left(\begin{matrix}
  0 & -i\\ i & 0\\
\end{matrix}\right), \qquad
\sigma_3=\left(\begin{matrix}
1 & 0\\ 0 & -1\\
\end{matrix}\right).$$
We follow here mainly the notation of Thaller's book \cite{Thaller}. 
We have chosen a system of units such that $\hbar=c=1$, and also such 
that the mass $m_e$ of the electron is normalized to 1.

The operator $D^0$ acts on $4$-spinors, i.e. functions $\psi\in 
\mathcal{H}:=L^2(\R^3,\C^4)$. It is self-adjoint on $\mathcal{H}$, 
with domain $H^1(\R^3,\C^4)$ and form domain $H^{1/2}(\R^3,\C^4)$. 
Moreover, it is defined to ensure
$$(D^0)^2=-\Delta+1.$$
The spectrum of $D^0$ is $(-\ii;-1]\cup [1;\ii)$. 
In what follows, the projector associated with the negative part of 
the spectrum of $D^0$ will be denoted by $P^0$:
$$P^0:=\chi_{(-\ii;0)}(D^0).$$
We then have
$$D^0P^0=P^0D^0=-\sqrt{1-\Delta}P^0=-P^0\sqrt{1-\Delta},$$
$$D^0(1-P^0)=(1-P^0)D^0=\sqrt{1-\Delta}(1-P^0)=(1-P^0)\sqrt{1-\Delta},$$
and
$$\mathcal{H}=\mathcal{H}_-^0\oplus\mathcal{H}_+^0,$$
where $\mathcal{H}_-^0:=P^0\mathcal{H}$ and 
$\mathcal{H}_+^0:=(1-P^0)\mathcal{H}$.

The fact that the 
spectrum of $D^0$ is not bounded from below is the source of many 
difficulties in Relativistic Quantum Mechanics. To explain why a free 
electron does not dissolve into the lower continuum, Dirac's idea \cite{D1,D}
was to postulate that in the absence of external field, the vacuum contains infinitely
many virtual electrons which completely fill up the negative part of the spectrum of
$D^0$. This \emph{Dirac Sea} should be seen as an infinite Slater determinant
$\Omega^0=\psi^0_1\wedge\cdots\wedge\psi^0_i\wedge\cdots$ where 
$(\psi^0_i)_{i\geq1}$ is an orthonormal basis of $\mathcal{H}_-^0$, 
whose density matrix is precisely
$$P^0=\sum_{i\geq1}|\psi^0_i\rangle\langle\psi^0_i|.$$
The projector $P^0$ is often called the \emph{bare vacuum} \cite{Chaix}.

\medskip

Let us now add an external coulomb field, created for instance by a system of
smeared nuclei. The density of protons in this system is a nonnegative function\footnote{However, we shall not limit to nonnegative $L^1$ 
densities $n$ in this paper, since the model we want to study is able to describe the 
vacuum interacting with both matter and antimatter.}
$n$ such that  $\int_{\R^3} n=Z$, the total number of protons
in the nuclei.
In our system of units, the external coulomb potential felt by the electrons is $-\alpha \phi$, where $\phi=n\ast\frac{1}{|\cdot|}$ and $\alpha$ is a small dimensionless coupling constant, usually called the Sommerfeld fine structure constant.
The Dirac operator with this external 
field is
\begin{equation}
D^{\alpha\phi}:=D^0-\alpha\phi.
\label{dirac}
\end{equation}

Dirac postulated that the charge of the bare vacuum is not measurable. Indeed, $P^0$ commutes with translations, so its density of charge must be constant, and cannot create any electric force.  However, in the presence of an external field, the virtual electrons react, by occupying the negative energy states of a new Dirac operator which does not commute with translations: \emph{the vacuum is polarized}. This 
polarization of the \emph{dressed vacuum}, which takes the form of a 
local density of charge, is measurable in practice.

The vacuum polarization plays a minor 
role in the calculation of the Lamb shift for the ordinary hydrogen 
atom (comparing to other electrodynamic phenomena), but it is important 
for high-$Z$ atoms \cite{MPS} and even plays a crucial role in muonic atoms 
\cite{FE,GRS}. It also explains the production of electron-positron pairs, observed experimentally in heavy ions collisions \cite{ABHTS,RMG,KS2,Seipp,FS}.

When the external field is 
not too strong, a good approximation (called the Furry picture \cite{Fu}) is to define the polarized 
vacuum as the 
projector$$P^{\alpha\varphi}:=\chi_{(-\ii;0)}(D^{\alpha\varphi}).$$
Note that in reality, the 
polarized vacuum modifies the electrostatic field, and the virtual 
electrons react to the corrected field. This remark naturally leads to a 
self-consistent equation for the dressed vacuum, and to a fixed-point 
iterative procedure for solving it. If one starts the procedure from 
$P^0$, the first iteration gives $P^{\alpha\varphi}$, and this explains why 
the Furry picture is a good approximation. But in practice, 
corrections to the Furry picture are necessary for high accuracy 
computations of electronic levels near heavy nuclei. These 
corrections can be interpreted as the second iteration in a Banach 
fixed-point algorithm (see, e.g., \cite[section 8.2]{MPS}).

In physics, self-consistent equations are usually derived as Euler-Lagrange 
equations of an energy functional. It is the case, for instance, in 
the nonrelativistic Hartree-Fock model \cite{HF}. Similarly, the 
self-consistent equation for the vacuum has a variational 
interpretation: it is satisfied by a minimizer of the 
\emph{Bogoliubov-Dirac-Fock} (BDF) energy functional. This functional was first 
introduced by Chaix and Iracane  \cite{Chaix}, as a possible cure to the 
fundamental problems associated with standard relativistic quantum 
chemistry calculations.

In these calculations, electrons near heavy 
nuclei are usually treated, in first approximation, by the Dirac-Fock 
model \cite{[swi]}, a variant of Hartree-Fock in which the kinetic energy operator 
$-\Delta/2$ is replaced by the free Dirac operator $D^0$. This approach gives
results that are in excellent agreement with experimental data \cite{[kim],[gra],[des],[lin-ro]}. When a higher accuracy is needed, the more sophisticated
multiconfiguration Dirac-Fock-model is used to take into account correlation effects \cite{[gor-al]},
and one can even compute the small corrections predicted by QED (vacuum polarization
and radiative corrections), using perturbation methods. However, the Dirac-Fock 
model suffers from an important defect: the corresponding energy is 
not bounded from below, contrary to the nonrelativistic Hartree-Fock case, and this leads to
important computational difficulties (see \cite{Chaix} for a discussion and detailed
references). From the mathematical viewpoint, one can 
prove that the Dirac-Fock functional has critical points which are solutions of the Dirac-Fock equations \cite{ES, P}, but these critical points have an infinite Morse index, and the rigorous definition of a ground 
state is delicate \cite{ES3,ES4}. 

The second problem with Dirac-Fock is its physical derivation: one would like to interpret this model as a variational approximation of Quantum Electrodynamics (QED), which is believed to be the exact theory. An interesting attempt in this direction has been made by Mittleman \cite{Mitt}, but it is not fully convincing. According to this author, the ground state of the Dirac-Fock model 
should be obtained by means of a max-min procedure applied to the no-photon QED Hamiltonian
${\bf H}_{\rm QED}$. 
In this procedure, the reference projector for the normal ordering of ${\bf H}_{\rm QED}$ is not 
fixed, and the vacuum polarization terms are neglected. Then 
one has to look for a projector which maximizes, in the Hartree-Fock 
approximation, the ground-state energy of the normal-ordered 
Hamiltonian. From a mathematical viewpoint, Mittleman's max-min 
principle has been investigated in the papers \cite{HF1,ES4,HF2,BES}. 
In the case of zero or one electron \cite{HF1,ES4}, it works very well and one shows that the projector $P^{\alpha\varphi}$ of the Furry picture is the optimal reference.
But it seems, from the counterexample given in \cite{BES}, that serious problems occur when there are several electrons.

In their work \cite{Chaix}, Chaix 
and Iracane derive their new mean-field model (which they call 
Bogoliubov-Dirac-Fock) from the no-photon QED Hamiltonian, 
normal-ordered with respect to a fixed reference: the free projector $P^0$. They keep the vacuum polarization terms, pointing out that they are ``necessary for the internal consistency of the relativistic mean-field theory and should therefore be taken into account in proper self-consistent calculations, independently of the magnitude of the physical effects" \cite[page 3813]{Chaix}. This allows them to obtain a bounded-below energy: the ground states are simply defined as minimizers and no max-min procedure is needed. A minimizer without charge constraint, if it exists, is a projector satisfying a self-consistent equation: it should be the 
negative spectral projector of the mean-field Hamiltonian generated 
by the nuclear charge density, corrected by a vacuum polarization 
effect. This self-consistent projector is the stable dressed vacuum. Now, if one restricts the BDF 
functional to the charge sector $-N$, and if one can find a 
minimizer, it will be solution of the Dirac-Fock equations for $N$ 
electrons, corrected by a vacuum polarization term \cite[section 4.2]{Chaix}. The Dirac-Fock model is thus interpreted by Chaix-Iracane as a {\it nonvariational} approximation of BDF. In other words, the Euler-Lagrange equations only differ by small terms, but the variational structure is completely different since the DF functional is strongly indefinite ({\it i.e.}, it is not bounded below and all its critical points have an infinite Morse index).

As we have seen, the Chaix-Iracane model has several advantages as compared to the 
standard Dirac-Fock model: it is more accurate (taking into account 
vacuum polarization effects), its physical derivation is more convincing, and the ground state solutions have a simple definition as minimizers of the BDF functional. The drawback is that it is
not easy to give a meaning to the quantities (energy of the vacuum, charge density of the vacuum) 
appearing in the BDF model. It is well known that there are divergent 
quantities in QED even after normal ordering, but Chaix and Iracane 
do not address this problem in their work. The first rigorous works on 
the BDF model are due to Chaix-Iracane-Lions \cite{CIL} and 
Bach-Barbaroux-Helffer-Siedentop \cite{HF1}. In particular, in \cite{HF1}, the
authors give a rigorous meaning to the BDF energy in the class 
of operators with trace, and show that it is bounded below if one 
fixes the reference for normal ordering. Then Bach {\it et al.} vary 
the reference for normal ordering and neglect the vacuum
polarization terms, which are experimentally small and mathematically 
divergent. This approximation is exactly the one made by Mittleman \cite{Mitt} in his formal derivation of the Dirac-Fock model.
In the present work, our approach is different: we keep $P^0$ as reference for normal ordering, we study the full Chaix-Iracane model of the dressed vacuum (without neglecting any divergent term), and we control the divergences thanks to a momentum cut-off. Nevertheless, the paper \cite{HF1} has been an important source of inspiration in our study: it contains very useful mathematical ideas and results, in particular the lower bound on the energy (see Theorem \ref{bounded} in the present paper, which is a mere rephrasing, in our framework, of this estimate).

\bigskip

Mathematically speaking, we shall say that a 
\emph{vacuum} is an orthogonal projector $P$ with the additional 
requirement that
$$Q=P-P^0\in\S_2(\mathcal{H})\;,$$
where $\S_2(\mathcal{H})$ denotes the 
class of Hilbert-Schmidt operators on $\mathcal{H}$. As 
explained in the Appendix, this condition guarantees that $P$ is the 
(unrenormalized) density matrix of a dressed vacuum in the 
electron-positron Fock space associated with the free projector $P^0$. 
This dressed vacuum may be seen (formally) as an infinite Slater 
determinant $\Omega=\psi_1\wedge\cdots\wedge\psi_i\wedge\cdots$ where 
$(\psi_i)_{i\geq1}$ is an orthonormal basis of 
$P\mathcal{H}$. Since the model takes the free 
vacuum as reference according to Dirac's ideas \cite{D1,D}, $Q$ is the true 
(renormalized) one-body density matrix of $\Omega$.
Following \cite{Chaix} (with notations from \cite{HF1}), the BDF energy of the dressed vacuum can be written (formally) as follows, as a function of its renormalized density matrix :
\begin{equation}
\label{formalenergy}
\mathcal{E}(Q)=\tr(D^0 
Q)-\alpha\int\rho_Q\phi+\frac{\alpha}{2}\int\!\!\!\int 
\frac{\rho_Q (x)\rho_Q (y)}{|x-y|}dx\,dy 
-\frac{\alpha}{2}\int\!\!\!\int \frac{|Q(x,y)|^2}{|x-y|}dx\,dy
\end{equation}
with $\rho_Q(x)=\tr_{\C^4} Q(x,x)$. By formal computations, Chaix 
and Iracane \cite[Section 4.2]{Chaix} show that the Euler-Lagrange equation of 
this functional is \begin{equation}
[P,D_{Q}]=0\;,
\label{equation_commutateur_D_Q}
\end{equation} where
\begin{equation}
D_{Q}:=D^{\alpha\phi} + \alpha\rho_Q\ast\frac{1}{|\cdot|} -\alpha 
\frac{Q(x,y)}{|x-y|},
\label{D_Q}
\end{equation} $Q=P-P^0$. For a  minimizer, the second order condition implies 
a more precise relation between $P$ and $D_Q$, which takes the form 
of a  \emph{fixed-point equation}:
\begin{equation}
\fbox{$\displaystyle 
P=\chi_{(-\ii;0)}(D_{Q})=\chi_{(-\ii;0)}\left(D^{\alpha\phi} + 
\alpha\rho_Q\ast\frac{1}{|\cdot|} -\alpha 
\frac{Q(x,y)}{|x-y|}\right).$}
\label{equation_P}
\end{equation}
Remark that if $\phi=0$ (no external potential), then $P^0$ is 
already a solution of this equation since $Q=0$ in this particular 
case.

The present paper contains the first mathematical study of a 
fixed-point algorithm for finding a solution of \eqref{equation_P}. Notice that the use of a fixed-point method to solve a 
self-consistent equation is very common in quantum chemistry and 
physics and that most of the numerical algorithms used in practice are based on this idea. For a mathematical existence result using the Schauder 
fixed-point  theorem, see the resolution of the Hartree equations in 
\cite{Wol}.  See also \cite{Cances}, where rigorous results are given 
on the convergence of standard Hartree-Fock iteration schemes. For 
the determination of a projector in no-photon QED, the fixed-point method has been used for the first time by Lieb and Siedentop \cite{LS}. Their goal was to replace $P^0$ by a new (self-consistent) projector commuting with translations, as reference for normal ordering in the absence of external field. We use the Banach fixed-point theorem as in \cite{LS}, but our physical model is very different, and the necessary estimates are much more delicate in our case. 

Of  course, we have to make an assumption on the external potential: it 
should have a certain regularity, and should not be too strong, otherwise we are not able to prove that the iteration method converges. If one is only interested in the 
existence of a minimizer, it is possible to remove the smallness 
assumption on the potential, but for this purpose the constructive fixed-point 
approach must  be replaced by a direct -- and non-constructive -- 
minimization argument \cite{HLS2}. The regularity assumption cannot be dropped: this is a well known phenomenon in QED when $P^0$ is chosen as reference for normal ordering (see e.g., \cite{K}). But this regularity is not really a restriction from the point of view of physics: point-like nuclei do not exist in nature.

\bigskip

In \cite{HF1},  the operator $D^0 Q$ is assumed to be trace class, so 
that  the expressions \eqref{formalenergy} and \eqref{D_Q} are 
well defined. Unfortunately, it turns out that, when $\phi$ is nonzero, $Q=P-P^0$ is 
\emph{never trace class} if $P$ is a solution of \eqref{equation_P}. Therefore no minimizer can exist in the trace class $\S_1(\mathcal{H})$ in the presence of an external field. So we must try to define the BDF energy and the self-consistent equation for operators which are not 
trace class, and this leads to several difficulties. 

A first problem 
occurs with the definition of $\tr(D^0 Q)$ in 
\eqref{formalenergy}. To solve it, we will have to extend the trace 
functional to a bigger class of compact operators, namely the operators with  ``$P^0$-trace" (see Section \ref{supertrace} 
below).

A second problem occurs with the definition of the density 
$\rho_Q$. For this reason, we introduce a momentum 
cut-off $\Lambda$, which means we 
replace the ambient space $\mathcal{H}$ 
by
$$\mathcal{H}_\Lambda:=\left\{f\in\mathcal{H},\ 
{\rm supp}(\widehat{f})\subset B(0,\Lambda)\right\}.$$
Since $D^0$ is a multiplication operator in Fourier space, 
$\mathcal{H}_\Lambda$ is invariant under $P^0$ and we keep the 
notation $P^0$ for the restricted operator.
With the cut-off, the integral kernel $Q(x,y)$ of 
$P-P^0\in\S_2(\mathcal{H}_\Lambda)$ becomes smooth for any dressed vacuum $P$, and
one can easily define $\rho_Q(x)=\tr_{\C^4}Q(x,x)$. Notice that, even with our ultraviolet cut-off, $Q=P-P^0$ is never trace class if $P$ is a solution of \eqref{equation_P} and if an external potential is present.
As we shall see it later on, our results will be valid under a technical condition 
of the form $\alpha\sqrt{\ln\Lambda}\leq C$ for some constant $C$. For a small $\alpha$, this 
leads to an extremely large $\Lambda$, which corresponds to scales 
that are far beyond the reach of experimental and theoretical physics 
at the present time. But our conditions do not allow to pass to the 
limit of an infinite cut-off. 

Note that if one expands 
the right-hand side of equation \eqref{equation_P} in powers of the 
small parameter $\alpha$, the first order term contains an expression 
which diverges logarithmically as $\Lambda$ goes to infinity. When the exchange term $Q(x,y)/|x-y|$ is neglected, a 
simple algebraic manipulation allows to rewrite {\it a posteriori} 
our cut-off version of equation \eqref{equation_P} in a 
renormalized form, with the divergent term removed, and the ``bare" 
constant $\alpha$ in front of the charge densities replaced by a 
smaller, ``dressed", coupling constant 
$$\alpha_{\rm dr} \simeq \frac{\alpha}{1 + \frac{2\alpha}{3\pi} \log\Lambda} $$
(details will be given in a forthcoming paper \cite{HLS2}).
The dressed constant is the observable one. Its experimental value is  $\alpha_{\rm dr}\simeq 1/137$. This kind of ``charge renormalization" associated with a momentum cut-off is standard in the physics literature (see, e.g., \cite[Equation (7.18)]{IZ}). With this interpretation, the 
limit case of an infinite cut-off appears as unphysical (it would 
correspond to $\alpha_{\rm dr}=0$, which means no more electrostatic 
interaction).

\begin{remark}\rm
In the Furry picture, that is to say when  $P=P^{\alpha\phi}$, it is known since the very beginning of QED \cite{D,H,FO,Ue,Se} that 
the density $\rho^{\alpha\phi}$ associated with $Q^{\alpha\phi}=P^{\alpha\phi}-P^0$ is \emph{never well-defined} if no ultraviolet cut-off is imposed. 
One possible regularization procedure \cite{FW,KS2,HS} is to remove the 
divergent part of $\rho^{\alpha\phi}$, which is (formally) 
proportional to the nuclear charge density $n$. This gives a 
renormalized density $\rho^{\alpha\phi}_{\mathrm{ren}}$ which can be defined without the 
help of a high momentum cut-off. This procedure has recently been clarified by Hainzl and Siedentop in \cite{HS}. Some 
interesting features of $\rho^{\alpha\phi}_{\mathrm{ren}}$, in the 
case of strong external fields, were obtained by Hainzl in 
\cite{Hainzl}. We do not want to give a precise definition of 
$\rho^{\alpha\phi}_{\mathrm{ren}}$ here and we refer the reader to 
\cite{HS,Hainzl}.

It would be tempting, 
instead of using a cut-off, to renormalize $\rho_Q$ {\it a priori} in 
equation $(\ref{equation_P})$, as in \cite{HS}. But we do not know 
how to solve the resulting renormalized equation $(\ref{equation_P})$ 
if no cut-off is made. Moreover, even if we could find a solution 
without momentum cut-off, its interpretation as a minimizer of the 
BDF energy would be unclear.
\end{remark} 
\bigskip

The paper is organized as follows. In the next section, we define the 
Bogoliubov-Dirac-Fock model and state our main results. For the sake 
of clarity, we have brought all the proofs together in Sections 
\ref{proofs1} and \ref{proofs2}. In the Appendix, we explain in our language, for the reader's convenience,
how the BDF energy is deduced from no-photon QED by Chaix-Iracane in \cite{Chaix}.

\bigskip

\noindent {\bf Acknowledgment: } {\it C.H. wishes to thank Heinz 
Siedentop for suggesting him the possibility of studying a 
self-consistent model of the polarized vacuum, during his time as 
post-doc at the LMU (Munich). The authors acknowledge support from 
the European Union's IHP network \emph{Analysis and Quantum} 
HPRN-CT-2002-00277.  E.S. acknowledges support from the Institut 
Universitaire de France.}

\section{Model and main results}
In this section, we study the Bogoliubov-Dirac-Fock model introduced 
in \cite{Chaix,Chaix_th}. Our system of notation is similar to 
\cite{HF1}, with the difference that we keep all the terms describing 
the vacuum polarization. This forces us to deal with operators which 
are not trace class, unlike \cite{HF1}.
\subsection{An extension of the trace functional}
\label{supertrace}
In order to give a meaning to the expression "$\tr(D^0 Q)$" even when $Q$ is not trace-class,
we need the notion of "$P^0$-trace".
In this section only, we work in an abstract Hilbert space $\EuFrak{h}$.
\begin{definition}
Let $P$ be a projector such that $P$ and $1-P$ have infinite rank, 
and $A\in\S_2(\EuFrak{h})$. We shall say that $A$ is 
\emph{$P$-trace class} if and only if 
$A_{++}:=(1-P)A(1-P)$ and $A_{--}:=PA P$ are trace class. Then we 
define the \emph{$P$-trace of $A$} by
$$\str_P(A):=\tr(A_{++})+\tr(A_{--}).$$
We denote by $\S_1^{P}(\EuFrak{h})$ the space of all Hilbert-Schmidt 
operators which are $P$-trace class.
\end{definition}

Notice that if $A$ is a trace class operator, then 
$A\in\S_1^{P}(\EuFrak{h})$ and $\tr(A)=\str_P(A)$ for any projector 
$P$. 
\begin{remark}\rm
In \cite[Section 5.7.2]{Thaller}, a similar definition in connection with supersymmetry is made and the name ``supertrace" is used.
\end{remark} 

The following result, whose proof is given in Section 
\ref{proofs1}, will be used repeatedly in the sequel.
\begin{lemma}\label{trace_blocks} Let $P$ and $P'$ be two projectors 
such that $P-P' \in \S_2(\EuFrak{h})$. Then $A$ is $P$-trace class 
if and only if it is $P'$-trace class, and in this case $\; \str_P(A)=\str_{P'}(A)$.
\end{lemma}
Another useful fact is that when $A$ is Hilbert-Schmidt and $A+P$ 
is a projector, then $A$ has a $P$-trace, as 
explained below:
\begin{lemma}
\label{lemma_blocks}
Let $P$ and $P'$ be two projectors on a Hilbert space, such that 
$P'-P$ is a Hilbert-Schmidt operator. Then $P'-P$ is $P$-trace class. Moreover, $\str_{P}(P'-P)$ is an integer which 
satisfies
$$\str_{P}(P'-P)=\tr\left( (P'-P)^{2n+1}\right)$$
for all $n\geq 1$, and $\str_{P}(P'-P)=0$ when $\norm{P'-P}_{\S_\infty} <1$.
\end{lemma}

In our framework, a consequence is that, for any vacuum $P$ such that $Q=P-P^0\in\S_2(\mathcal{H}_\Lambda)$,  $\str_{P^0}(Q)$
is an integer which can be interpreted as the charge of the dressed vacuum $P$ (see the Appendix for comments in this direction).
When $P$ solves the self-consistent 
equation \eqref{equation_P} and $\phi$ is not too strong, we will see that $P$ is close to $P^0$,
so that its charge will be zero, 
according to the lemma.

\subsection{The Bogoliubov-Dirac-Fock model}
\label{bdfmodel}

As in \cite{HF1}, we are going to extend the BDF energy to a convex set of compact operators, which can be interpreted as one-particle density matrices of quasi-free states. This kind of extension is standard for mean-field models 
depending only on the one-body density matrix (see \cite{Lieb,Bach,quasi}).

\medskip

In the whole paper, we 
assume that the nuclear charge density $n=-\Delta \phi/4\pi$ belongs to 
the Hilbert space
$$\EuFrak{C}=\left\{f \in L^2(\R^3,\R),\ 
D(f,f) <\ii\right\},$$
where
$$D(f,g)=4\pi\int\frac{\overline{\widehat{f}(k)}{\widehat{g}(k)}}{|k|^2}dk.$$
We will choose the following Hilbert norm on $\EuFrak{C}$ :
$$\normr{f}:=\left(\int\frac{1+|k|^2}{|k|^2}|\widehat{f}(k)|^2\,dk\right)^{1/2}\;.$$
The Bogoliubov-Dirac-Fock energy is defined by
\begin{equation}
\label{energy}
\mathcal{E}(\Gamma)=\str_{P^0}(D^0 
\Gamma)-\alpha D({\rho}_\Gamma,n) +\frac{\alpha}{2}D(\rho_\Gamma, \rho_\Gamma)
-\frac{\alpha}{2}\int\!\!\!\int \frac{|\Gamma(x,y)|^2}{|x-y|}dx\,dy
\end{equation}
on the set
\begin{equation}
\mathcal{G}_\Lambda:=\left\{\Gamma\in 
\S_1^{P^0}(\mathcal{H}_\Lambda)\ |\  -P^0\leq\Gamma\leq 1-P^0,\ 
\rho_\Gamma\in\EuFrak{C}\right\}.
\label{BLambda}
\end{equation} In (\ref{energy}), 
$$\widehat{\rho}_\Gamma(k)=\frac{1}{(2\pi)^{3/2}}\int_{|p|\leq\Lambda} 
\Tr{\widehat{\Gamma}(p+k/2,p-k/2)}dp$$is the Fourier transform of 
the charge density $\rho_\Gamma$, which, formally, is the diagonal of 
$\Gamma\in\S_2(\mathcal{H}_\Lambda)$, as explained in the 
Introduction. Thanks to the momentum cut-off, $\widehat{\rho_\Gamma}$ is compactly supported, so that 
$\widehat{\rho}_\Gamma \in L^1$, hence
\begin{equation}
\label{rho}
\rho_\Gamma(x)=\Tr{\Gamma(x,x)}=\frac{1}{(2\pi)^3}\int\!\!\!\int_{|p|,|q|\leq\Lambda} 
\Tr{\widehat{\Gamma}(p,q)}e^{ix(p-q)}dp\,dq\,.
\end{equation}
Clearly, the function $\Gamma\in \S_2(\mathcal{H}_\Lambda)\mapsto 
\rho_\Gamma\in {\cal C}^0_0\cap L^2(\R^3)$ is 
continuous.
Notice that if for instance $f,g\in H^1(\R^3,\R)$, then the electrostatic energy is simply
$$D(f,g)=\iint_{\R^3\times\R^3}\frac{f(x){g(y)}}{|x-y|}dx\,dy,$$
but for functions in $\EuFrak{C}$,  it does not necessarily have a meaning as a Lebesgue double integral in direct 
space. 

\medskip

Note that the set $\mathcal{G}_\Lambda$ is 
convex, and that the elements of $P^0+\mathcal{G}_\Lambda$ are not 
necessarily projectors. In fact, it is an easy exercise to show that 
an element of $\mathcal{G}_\Lambda$ is extremal if and only if it is 
of the form $P-P^0$, with $P$ a projector. It then follows from Lemma 
\ref{lemma_blocks} that the set of all extremal points of 
$\mathcal{G}_\Lambda$ coincides with 
$\mathcal{Q}_\Lambda=\mathcal{P}_\Lambda-P^0$, 
where$$\mathcal{P}_\Lambda=\left\{P \mbox{ orth. projector} \ |\ 
Q=P-P^0\in\S_2(\mathcal{H}_\Lambda),\ \rho_Q\in{\EuFrak{C}}
\right\}.$$
It will turn out that, under some assumptions on 
$\alpha$, $\Lambda$ and $n$, the BDF functional has a unique 
minimizer on $\mathcal{G}_\Lambda$ which is extremal. As a 
consequence,$$\inf\{{\cal E}(P-P^0),\ 
P\in\mathcal{P}_\Lambda\}=\inf\{{\cal E}(\Gamma),\ 
\Gamma\in\mathcal{G}_\Lambda\}\;. $$

In the next subsection, we give necessary and sufficient conditions 
satisfied by a minimizer of ${\cal E}$.

\subsection{Study of the BDF energy}
We first state the following result, which is an easy translation, in 
our framework, of the stability estimate proved by  Bach {\it et al} 
\cite{HF1} (see also \cite{CIL}):

\begin{thm}\label{bounded} Let be $n\in \EuFrak{C}$. Then
\begin{enumerate}
\item $\mathcal{E}$ is well-defined on $\mathcal{G}_\Lambda$;
\item if $0\leq \alpha\leq \frac{4}{\pi}$, then
\begin{equation}
\forall \Gamma\in\mathcal{G}_\Lambda,\qquad 
\mathcal{E}(\Gamma)+\frac{\alpha}{2} D(n,n)\geq 
0
\label{bound_energy_below}
\end{equation}
and therefore $\mathcal{E}$ is bounded from below on 
$\mathcal{G}_\Lambda$, independently of $\Lambda$;
\item if $0\leq \alpha\leq \frac{4}{\pi}$ and $n=0$, then 
$\mathcal{E}$ is non-negative on $\mathcal{G}_\Lambda$ 
\cite{HF1,CIL}, 0 being the unique minimizer.
\end{enumerate}
\end{thm}

\begin{remark}\rm
Note that the result is optimal in the sense that the functional 
becomes unbounded from below when $n=0$ if $\alpha > 4/\pi$, as shown 
in \cite{CIL} and \cite{HRS}.
\end{remark} 

\begin{remark}\rm
Since $\alpha D(n,n)/2$ in 
\eqref{bound_energy_below} is the electrostatic energy of the field 
created by $n$, \eqref{bound_energy_below} means that the 
total energy of the system is nonnegative.
\end{remark} 


\noindent{\it Proof of Theorem \ref{bounded}.}
We only explain here why $\mathcal{E}$ is well defined on $\mathcal{G}_\Lambda$, the rest being identical to the proof of Theorem 1 in \cite{HF1} (see  \cite[Eq. $(18)$-$(19)$]{HF1}).

If $\Gamma\in \S_1^{P^0}(\mathcal{H}_\Lambda)$, then we have
$$P^0D^0\Gamma P^0=D^0P^0\Gamma P^0=D^0\Gamma_{++}\in \S_1(\mathcal{H}_\Lambda)$$
since $P^0$ commutes with $D^0$ and $|D^0|\leq \sqrt{1+\Lambda^2}$. With a similar argument for $1-P^0$, we obtain that $D^0\Gamma\in \S_1^{P^0}(\mathcal{H}_\Lambda)$. Therefore, $\str_{P^0}(D^0\Gamma)$ is well-defined and 
\begin{equation}
\label{kinetic}
\str_{P^0}(D^0\Gamma)=\tr(D^0\Gamma_{++})+\tr(D^0\Gamma_{--})=\tr(|D^0|\Gamma_{++})-\tr(|D^0|\Gamma_{--})
\end{equation} 
(notice that, due to the constraint $-P^0\leq \Gamma\leq1-P^0$, one has $\Gamma_{++}\geq0$ and $\Gamma_{--}\leq 0$).
On the other hand we have by Kato's inequality
$$\int\!\!\!\int\frac{|\Gamma(x,y)|^2}{|x-y|}dx\,dy\leq \frac{\pi}{2}\tr(|D^0|\Gamma^2)$$
showing that this last term is well-defined since $|D^0|$ is bounded on $\mathcal{H}_\Lambda$ and $\Gamma\in \S_2(\mathcal{H}_\Lambda)$. \qed

\bigskip

We are interested in minimizers of the BDF functional, and we expect 
them to be in the class ${\cal Q}_\Lambda={\cal P}_\Lambda-P^0$. This 
leads to the following definition
\begin{definition}
\label{stability_def}
We say that a projector $P$ is a \emph{BDF-stable vacuum} if and only 
if $P-P^0$ is a minimizer of $\mathcal{E}$ on $\mathcal{G}_\Lambda$.
\end{definition}

When there is no external potential, $P^0$ is the unique BDF-stable vacuum  \cite{CIL,HF1}, which corresponds to 
Dirac's ideas.  But if we consider a non-vanishing external potential 
$\phi= n\ast  \frac 1{|\cdot|}$, then $P^0$ obviously cannot be 
BDF-stable, since it is easy to create a state $-P^0 \leq \gamma \leq 
1 - P^0$ such that $\mathcal{E}(\gamma) < 0 =\mathcal{E}(0)$. This 
means that the vacuum is necessarily polarized. 

More precisely, one 
can easily derive necessary conditions satisfied by a BDF-stable vacuum $P$. To 
this end, a perturbation of the form $Q+\gamma=P-P^0+\gamma$, with $\gamma\in\S_1(\mathcal{H}_\Lambda)$ such that $-P\leq\gamma\leq1-P$ is considered in Chaix-Iracane \cite[formula $(4.8)$]{Chaix}, and the energy ${\cal E}(Q+\gamma)$ is expanded to get
\begin{equation}\label{separation}
\mathcal{E}(Q+\gamma)   =  \tr\left(D_Q\gamma\right) 
+\frac{\alpha}{2}D(\rho_\gamma,\rho_\gamma)-\frac{\alpha}{2}\int\!\!\!\int\frac{|\gamma(x,y)|^2}{|x-y|}dx\,dy+\mathcal{E}(Q),
\end{equation}
a formula which is valid when $\gamma\in\S_1(\mathcal{H}_\Lambda)$, the operator $D_Q$ being defined in \eqref{D_Q}. 

\bigskip

\begin{remark}\rm
In \cite[Formula $(21)$]{HF1} and \cite{HF2}, the polarization potentials appearing 
in $D_Q$ and the energy of the vacuum $\mathcal{E}(Q)$ were neglected 
by the authors who used the following functional
\begin{equation}
\tilde{\mathcal{E}}_P(\gamma)  = 
\tr\left(D^{\alpha\phi}\gamma\right) 
+\frac{\alpha}{2}D(\rho_\gamma,\rho_\gamma)-\frac{\alpha}{2}\int\!\!\!\int\frac{|\gamma(x,y)|^2}{|x-y|}dx\,dy,
\label{BBHS_energy}
\end{equation}
with the constraints $\gamma\in\S_1$ and $-P\leq\gamma\leq1-P$ (and even $P\gamma(1-P)=0$ in \cite{HF2}). Then a procedure taking 
the form $\sup_P\inf_{-P\leq\gamma\leq1-P}\tilde{\mathcal{E}}_P(\gamma)$, related to Mittleman's work \cite{Mitt}, was 
considered in \cite{HF1}. For the case of the vacuum (no constraint on the trace of $\gamma$), the 
solution is the Furry picture $P=P^{\alpha\phi}$ with $\gamma=0$, as shown in 
\cite{HF1}. We refer the reader to \cite[page 3809]{Chaix} and 
\cite{ES4,BES,HF2} for comments and results concerning Mittleman's max-min 
in the case of $N$ electrons (which corresponds to the additional constraint $\tr(\gamma)=N$).
\end{remark} 

\bigskip

>From formula \eqref{separation}, it can be seen that a BDF-stable vacuum must satisfy the fixed-point equation \eqref{equation_P}. The converse is also true under some assumptions:

\begin{thm}[BDF-Stability]\label{stability}
Let be $P\in\mathcal{P}_\Lambda$ and $n\in \EuFrak{C}$. We assume 
that there exists a positive constant $d$ such that
\begin{equation}
\label{estim_d}
d|D_Q|\geq |D^0|\ \mbox{ with }\ \alpha d\frac{\pi}{4}\leq 1,
\end{equation}
where $D_Q$ is defined in $(\ref{D_Q})$.
Then, the following assertions are equivalent
\begin{enumerate}
\item $P$ fulfills the equation
\begin{equation}
\label{equation_thm}
P=\chi_{(-\ii;0)}(D_{Q})=\chi_{(-\ii;0)}\left(D^{\alpha\phi} + 
\alpha\rho_Q\ast\frac{1}{|\cdot|} -\alpha \frac{Q(x,y)}{|x-y|}\right).
\end{equation}
\item $P$ is the unique BDF-stable vacuum, i.e. $P-P^0$ is the unique global minimizer of $\mathcal{E}$ 
on $\mathcal{G}_\Lambda$.
\end{enumerate}
\end{thm}
The proof of this theorem will be given in Section 3. Some arguments are directly inspired of \cite{HF1}.

\begin{remark}\rm
One can try to go further: if we translate the ideas of Chaix 
and Iracane in our language, the ground state of a molecule consisting of 
nuclei with total charge density $n$, surrounded by a cloud of $N$ 
electrons, should solve the following constrained minimization 
problem
$$\min\{{\cal E}(\Gamma),\ \Gamma\in\mathcal{G}_\Lambda, \ 
\str_{P^0}(\Gamma)=N\}, $$
with $N\in\N\setminus\{0\}$. 
If this minimization problem has a 
solution $Q$, it will solve a self-consistent equation of the 
form 
$$Q=\chi_{(-\ii;\mu)}(D_{Q})-P^0,$$
where $\mu\in(-1;1)$ is a Lagrange 
multiplier associated with the charge constraint, and interpreted as 
a chemical potential. For a not too strong external field $\phi$, it should be possible to prove that $\mu\in(0;1)$ and that the vacuum $\Pi=\chi_{(-\ii;0)}(D_{Q})$ stays neutral, which means $\str_{P^0}(\Pi-P^0)=0$. Therefore, we could split $P=Q+P^0=\chi_{(-\ii;\mu)}(D_{Q})$ in the 
form$$P=\Pi+\sum_{k=1}^{N}|\psi_k\rangle\langle\psi_k 
|.$$
The mono-electronic wave 
functions $\psi_k$ would be solutions of the Dirac-Fock equations 
(with high momentum cut-off), perturbed by vacuum polarization 
terms:$$D_Q\psi_k=\varepsilon_k \psi_k\;,\quad 0<\varepsilon_k < 
1\;,\quad 1\leq k\leq N\;.$$It is our goal to study this constrained 
variational problem in the near future. The present work, which deals 
with the unconstrained case, is a first step in this direction.
\end{remark}

\subsection{Existence of a BDF-stable vacuum}
We may now state our main Theorem. We recall that the norm on 
${\EuFrak{C}}$ is
$$\normr{f}:=\left(\int\frac{1+|k|^2}{|k|^2}|\widehat{f}(k)|^2\,dk\right)^{1/2}\;.$$

\begin{thm}[Existence of a BDF-stable vacuum] \label{existence} Let 
be $n\in\EuFrak{C}$ and $b\in(0;1)$. Then for all 
$\Lambda$ and $\alpha$ such that
\begin{equation}
2\sqrt{\pi}\alpha  \normr{n}\leq b \ \ \mbox{ and }\ \ \alpha\leq 
\alpha_b(\Lambda)
\label{conditions_thm}
\end{equation} 
where
$$\alpha_b(\Lambda)\sim_{\Lambda\to\ii}\frac{C(1-b)}{\sqrt{\log\Lambda}},$$
there exists a unique BDF-stable vacuum $P$, which is a solution of
\begin{equation}
 P=\chi_{(-\ii;0)}\left(D_Q\right)=\chi_{(-\ii;0)}\left(D^0+\alpha 
(\rho_Q-n)\ast\frac{1}{|\cdot|}-\alpha\frac{Q(x,y)}{|x-y|}\right)
\label{equation_P_thm}
\end{equation} 
with $Q=P-P^0$.  Moreover, we have $\str_{P^0}(Q)=0$.
\end{thm}

\begin{remark}\rm
The first constraint $2\sqrt{\pi}\alpha  \normr{n}\leq b$ means that the external field is not too strong. It explains why a neutral polarized vacuum is obtained (since $\str_{P^0}(Q)=0$). In our proof, this constraint on the external field is necessary for the fixed point algorithm to converge. The second constraint $\alpha\leq 
\alpha_b(\Lambda)$, which essentially reduces to $\alpha\sqrt{\log\Lambda}\lesssim C(1-b)$, is a technical condition due to our choice of norms, but we were unable to drop it. It disappears if the exchange term $\frac{\alpha}{2}\iint\frac{|Q(x,y)|^2}{|x-y|}dx\,dy$ is neglected in the energy, as
can be seen from the proof. A precise definition of the constant $C$ appearing in this result is 
given in the proof. 
\end{remark}

\begin{remark}\rm
There is an interesting symmetry property of the solutions of \eqref{equation_P_thm} when $n$ is replaced by $-n$. Namely, if $P$ is a solution of \eqref{equation_P_thm} with external density $n$, then $P'=Q'+P^0$ is a solution of \eqref{equation_P_thm} with external density $-n$, where $Q'=-CQC^{-1}$, $C$ being the charge conjugation operator \cite[page 14]{Thaller}. The two dressed vacua $P$ and $P'$ have the same BDF energies and satisfy $\rho_{Q'}=-\rho_Q$, as suggested by the intuition. For this symmetry between matter and antimatter to be true, it is essential to have the Fermi level at $0$ and not at $-1$ (see, e.g., the comments of \cite[page 197]{SS} about this fact).
\end{remark} 

\subsection{Idea of the proof of Theorem \ref{existence}: the fixed-point algorithm}

We end this section with a brief description of our fixed-point algorithm, used in the proof of Theorem \ref{existence}, and which could be useful for practical computations.
\medskip

A natural scheme for solving \eqref{equation_P_thm} would be to construct a sequence $(Q_j)_{j\geq0}\subset\mathcal{G}_\Lambda$ by taking $Q_0=0$ and
\begin{equation}
Q_{j+1}=\chi_{(-\ii;0)}\left(D^0+\alpha 
(\rho_{Q_j}-n)\ast\frac{1}{|\cdot|}-\alpha\frac{Q_j(x,y)}{|x-y|}\right)-P^0.
\label{scheme1}
\end{equation}
Expanding this expression in powers of $\alpha$, and considering the total density
$$\rho'_{Q_{j}}:=\rho_{_{Q_j}}-n,$$
one can write the following recursion formula in Fourier space:
\begin{equation}
\widehat{\rho'_{Q_{j+1}}}(k)=-\widehat{n}(k)-\alpha B_\Lambda(k)\widehat{\rho'_{Q_{j}}}(k)+\alpha\,\widehat{\rho_{1,0}\scriptstyle{(Q_j)}}(k)+
\sum_{n=2}^\ii \alpha^n\,\widehat{\rho_n \scriptstyle{(\!Q_j ,\,\rho'_{\scriptscriptstyle{\!Q_j}}\!)}\,}(k)\;.
\label{equation_rho_algo1}
\end{equation}
The notations $B_\Lambda\,$, $\rho_{1,0}$ and $\rho_n$ are defined precisely in Section \ref{prelim} (see the subsections \ref{Cauchy_formula} and \ref{first_order_density}). The important point is that $B_\Lambda(k)$ is a positive function which diverges logarithmically as $\Lambda\to\ii$ for any fixed $k$, whereas the other terms stay bounded. From (\ref{equation_rho_algo1}) we thus see that the scheme (\ref{scheme1}) would converge under a condition of the form $\alpha\log\Lambda\leq C$.

To improve this condition, we use a better algorithm in our proof. Our modified scheme consists in defining a sequence of pairs $(Q_j,\rho'_j)_{j\geq0}\subset\mathcal{G}_\Lambda\times\EuFrak{C}$ such that $(Q_0,\rho'_0)=(0,-n)$ and
\begin{equation}
\left\{\begin{array}{l}
\displaystyle Q_{j+1}=\chi_{(-\ii;0)}\left(D^0+\alpha 
\rho'_{j}\ast\frac{1}{|\cdot|}-\alpha\frac{Q_j(x,y)}{|x-y|}\right)-P^0\\
\displaystyle \rho'_{j+1}=L(\alpha,\Lambda) \rho'_{Q_{j+1}}+\big(1-L(\alpha,\Lambda)\big)\rho'_j
\end{array} \right.
\label{scheme2}
\end{equation} 
where $\rho'_{Q_j}(x):=\rho_{_{Q_j}}(x)-n(x)=\tr_{\C^4}Q_j(x,x) - n(x)$, and $L(\alpha,\Lambda)$ is the linear operator which, in the Fourier domain, is just the multiplication by the function $(1+\alpha B_\Lambda(k))^{-1}$. The second equation in the iteration scheme (\ref{scheme2}) can be written in the form
\begin{equation}
\widehat{\rho'_{j+1}}(k)=(1+\alpha B_\Lambda(k))^{-1}\bigl[-\widehat{n}(k)+\alpha\,\widehat{\rho_{1,0}\scriptstyle{(Q_j)}}(k)+
\sum_{n=2}^\ii \alpha^n\widehat{\,\rho_n \scriptscriptstyle{(\!Q_j,\,\rho'_j\!)}\,}(k)\,\bigr ]
\label{equation_rho_algo2}
\end{equation} 
The divergent term now only appears in the denominator. So one expects a much better convergence. In the proof of Theorem \ref{existence}, we show the convergence of the algorithm (\ref{scheme2}) under the conditions \eqref{conditions_thm} but we believe that it converges independently of the cut-off $\Lambda$. 

It can be seen from our proof that this holds when the exchange term $\alpha\frac{Q_j(x,y)}{|x-y|}$ is neglected in (\ref{scheme2}). In this case, the algorithm converges independently of $\Lambda$ to the solution of a reduced fixed-point problem (without exchange term), which is the unique minimizer on $\mathcal{G}_\Lambda$ of the convex functional
$$\mathcal{E}_{\rm red}(\Gamma)=\str_{P^0}(D^0 
\Gamma)-\alpha D({\rho}_\Gamma,n)+\frac{\alpha}{2}D(\rho_\Gamma, \rho_\Gamma)\; .$$

\section{Proof of Theorem \ref{stability}}
\label{proofs1}
In this section, we prove Theorem \ref{stability}. To this end, we 
first need to prove Lemmas \ref{trace_blocks} and \ref{lemma_blocks}.

\subsection{Proof of Lemmas  \ref{trace_blocks} and \ref{lemma_blocks}}  
\noindent {\it Proof of Lemma 
\ref{trace_blocks}.}
Let be $P$ and $P'$ two projectors such that 
$P'-P\in\S_2(\EuFrak{h})$, and a Hilbert-Schmidt operator $A$ which 
is $P$-trace class. This means that $PAP$ and 
$(1-P)A(1-P)$ are trace class.

Let us first show that $P'AP'$ is trace class. To this end, we write
\begin{eqnarray}
P'AP' & = & (P'-P+P)A(P'-P+P)\nonumber\\
  & = & (P'-P)A(P'-P) + (P'-P)AP + PA(P'-P) + PAP.\qquad \label{chgt_proj}
\end{eqnarray}
This shows that $P'AP'$ is trace class since the last term is in 
$\S_1$ by assumption, $P'-P$ and $A$ are in $\S_2$, and $P$ is 
bounded. The same computation shows that $(1-P')A(1-P')$ is 
trace class.

We now compute
\begin{eqnarray*}
\tr[P'AP'] & = & \tr[(P'-P)A(P'-P)] +\tr[(P'-P)AP] + \tr[PA(P'-P)] + \tr[PAP]\\
  & = & \tr[A\left((P'-P)(P'-P)+P(P'-P)+(P'-P)P\right)] + \tr[PAP]\\
  & = & \tr[A(P'-P)] + \tr[PAP],
\end{eqnarray*}
where we have used the formula $\tr(AB)=\tr(BA)$, valid for 
$A,B\in\S_2$. The same computation gives
$$\tr[P'_+AP'_+]  = \tr[A(P'_+-P_+)] + \tr[P_+AP_+] = -\tr[A(P'-P)] + 
\tr[P_+AP_+],$$
where we have used the notation $P_+=1-P$ and $P'_+=1-P'$.
Summing this two results, we obtain the formula 
$\str_{P}[A]=\str_{P'}[A]$. \qed
   
\medskip 

\noindent {\it Proof of Lemma 
\ref{lemma_blocks}.}  We introduce $B=P'-P$. We have $B^2=P'-P'P-PP'+P=(1-P)B(1-P)-PBP$. 
This implies that $(1-P)B(1-P)$ and $-PBP$ are non-negative 
trace class operators.
We now use the proof of \cite[Theorem 4.1]{index}. Since $B\in\S_2$, 
we infer $B^3\in\S_1$ and so $(P',P)$ is a Fredholm pair, in the 
language of \cite{index}. Therefore, $\tr(B^3)$ is an integer and 
satisfies $\tr(B^3)=\tr(B^{2n+1})$ for all $n\geq 1$. Now we have
$$B^3 = B^2P'- B^2P = P'BP'+PBP.$$
Applying this result to $1-P'$ and $1-P$, we find
$$B^3 = (1-P')B(1-P')+(1-P)B(1-P).$$
Summing this two identities, we obtain by Lemma \ref{trace_blocks}
$$2\tr(B^3)=\str_{P'}(B)+\str_{P}(B)=2\str_{P}(B).$$
This shows that $\str_{P}(B)$ indeed equals the index of the pair of 
projectors $(P',P)$ defined in \cite{index}, an integer which 
vanishes when $\norm{P'-P}_{\S_\infty}<1$ by the results of \cite{index}.\qed

\subsection{Preliminaries}
To prove Theorem \ref{stability}, we also need the following
\begin{lemma}
\label{injections}
Assume that $\phi=\rho\ast\frac{1}{|\cdot|}$ for some $\rho\in\EuFrak{C}$. Then
$$\norm{\nabla\phi}_{H^{1}}=4\pi\normr{\rho},\qquad 
\norm{\phi}_{L^\ii}\leq C_\ii4\pi\normr{\rho},\qquad 
\norm{\phi}_{L^6}\leq C_64\pi\normr{\rho}$$
where $C_\ii:=\frac{1}{2\pi^{1/2}}$ and $C_6$ is the Sobolev constant 
for the inequality $\norm{u}_{L^6(\R^3)}\leq C_6\norm{\nabla 
u}_{L^2(\R^3)}$.
\end{lemma}

\noindent {\it Proof.} We have
$$\int_{\R^3}\frac{1+|k|^2}{|k|^2}|\widehat{\rho}(k)|^2\,dk=\frac{1}{(4\pi)^2}\int_{\R^3}|k|^2(1+|k|^2)|\widehat{\phi}(k)|^2\,dk,$$
and so
$$\norm{\phi}_{L^\ii}\leq 
\frac{1}{(2\pi)^{3/2}}\norm{\hat{\phi}}_{L^1}\leq 
\frac{1}{(2\pi)^{3/2}}\left(\int_{\R^3}\frac{dk}{|k|^2(1+|k|^2)}\right)^{1/2}4\pi\normr{\rho}.$$
The rest is easily obtained by the Sobolev inequalities. \qed

\begin{lemma}
\label{DQ_bounded}
Let $P$ be a projector in $\mathcal{P}_\Lambda$ and $Q=P-P^0$. Then 
$D_Q$ is bounded.
\end{lemma}

\noindent {\it Proof.} Due to the cut-off in Fourier space, $D^0$ is bounded 
on $\mathcal{H}_\Lambda$. On the other hand, if 
$\phi=\rho\ast\frac{1}{|\cdot|}$ for some $\rho\in\EuFrak{C}$, then 
$\phi\in L^\ii$ by Lemma \ref{injections} and so this is also a 
bounded operator. Let us now denote $R(x,y)=\frac{Q(x,y)}{|x-y|}$. We 
then have
$$|Rf(x)|^2 = \left|\int_{\R^3}\frac{Q(x,y)f(y)}{|x-y|}\,dy\right|^2 
\leq \left(\int_{\R^3}\frac{|Q(x,y)|^2}{|x-y|}\,dy\right)\times 
\left(\int_{\R^3}\frac{|f(y)|^2}{|x-y|}\,dy\right)$$
and since, by Kato's inequality,
$$\int_{\R^3}\frac{|f(y)|^2}{|x-y|}\,dy\leq \frac{\pi}{2}\pscal{f,|D^0|f}$$
\begin{equation}
\label{Hardy}
\int\!\!\!\int_{\R^6}\frac{|Q(x,y)|^2}{|x-y|}\,dx\,dy\leq 
\frac{\pi}{2}\tr(|D^0|\,Q^2)
\end{equation}
this shows that $R\leq C\,|D^0|^{1/2}$ and so $R$ is bounded. \qed

\begin{lemma}
\label{lemme_egalite}
Let $P$ be a projector in $\mathcal{P}_\Lambda$ and $Q=P-P^0$. Then 
$D_Q\Gamma\in\S_1^{P_0}(\mathcal{H}_\Lambda)$ for all $\Gamma\in 
\mathcal{G}_\Lambda$ and we have
$$\str_{P^0}(D^0\Gamma)+\alpha D(\rho_Q-n,\rho_{\Gamma})-\alpha\tr\left(\frac{Q(x,y)}{|x-y|}\Gamma\right)=\str_{P^0}(D_Q\Gamma).$$
\end{lemma}

\noindent {\it Proof.} Remark that $R\Gamma=\frac{Q(x,y)}{|x-y|}\Gamma$ is 
trace class, since $\frac{Q(x,y)}{|x-y|^{1/2}}$ and 
$\frac{\Gamma(x,y)}{|x-y|^{1/2}}$ are in $\S_2$ by \eqref{Hardy}. Let 
us now define 
$D=D^{-\alpha(\rho_Q-n)\ast\frac{1}{|\cdot|}}=D^0+\alpha(\rho_Q-n)\ast\frac{1}{|\cdot|}$ 
and $P'=\chi_{(-\ii;0)}(D)$. By the result of Klaus-Scharf 
\cite{Scharf1} (see also \cite{HeSi} and the proof of Theorem 
\ref{existence}), it is known that 
$P'-P^0\in\S_2(\mathcal{H}_\Lambda)$. Thus $P'D\Gamma P'=DP'\Gamma 
P'\in\S_1(\mathcal{H}_\Lambda)$ since $\Gamma\in 
\S_1^{P_0}=\S_1^{P'}$ by Lemma \ref{trace_blocks}, and $D$ is bounded 
by the proof of Lemma \ref{DQ_bounded}. Therefore, 
$D_Q\Gamma=D\Gamma+R\Gamma$ is in $\S_1^{P_0}$.

To show the expected equality, we prove
\begin{equation}
\label{egalite}
\str_{P^0}(D^0\Gamma)+\alpha D(\rho'_Q,\rho_{\Gamma})=\str_{P^0}(D\Gamma)
\end{equation}
where $\rho'_Q=\rho_Q-n\in\EuFrak{C}$. This will end the proof since the 
other term is trace class. The general idea of the proof is to 
approximate $\Gamma$ by a trace class operator for which this 
equality is true, and to pass to the limit. However, the behaviour of 
the associated density in the space $\EuFrak{C}$ is not obvious and 
to overcome this difficulty, we shall also approximate the density 
$\rho'_Q$ to obtain a potential in $L^2(\R^3)$.
We thus start by choosing a sequence $\rho_j$ which converges as 
$j\to+\ii$ to $\rho'_Q$ in $\EuFrak{C}$, such that 
$\phi_j=\rho_j\ast\frac{1}{|\cdot|}$ is in $L^2(\R^3)$. We can choose 
for instance 
$\widehat{\rho_j}(k)=\widehat{\rho'_Q}(k)\chi_{_{(|k|\geq 1/j)}}$. We 
now show
\begin{equation}
\label{egalite2}
\str_{P^0}(D^0\Gamma)+\alpha 
\int_{\R^3}\rho_{\Gamma}\phi_j=\str_{P^0}(D_j\Gamma)
\end{equation}
for all $\Gamma\in\mathcal{G}_\Lambda$, and where 
$D_j=D^0-\alpha\phi_j$. To this end, we may find a sequence 
$\Gamma^n_{+-}$ of finite rank operator which converges to 
$\Gamma_{+-}=(1-P^0)\Gamma P^0$ in $\S_2$. Then
$$\Gamma^n:=\left(\begin{matrix}
\Gamma_{++} & \Gamma_{+-}^n\\
(\Gamma_{+-}^n)^* & \Gamma_{--}\\
\end{matrix}\right) $$
converges to $\Gamma$ in $\S_2$. Since $\Gamma^n\in\S_1$ for all 
$n\geq 0$, we have
\begin{equation}
\label{egalite3}
\str_{P^0}(D^0\Gamma^n)+\alpha 
\int_{\R^3}\rho_{\Gamma^n}\phi_j=\str_{P^0}(D_j\Gamma^n).
\end{equation}
By \eqref{rho}, the function $Q\in\S_2(\mathcal{H}_\Lambda)\mapsto 
\rho_Q\in L^2(\R^3)$ is continuous. Therefore, 
$\rho_{\Gamma^n}\to\rho_\Gamma$ in $L^2(\R^3)$. Since $\phi_j\in 
L^2(\R^3)$, we may now pass to the limit in \eqref{egalite3} and 
obtain
$$\lim_{n\to\ii}\left( \str_{P^0}(D^0\Gamma^n)+\alpha 
\int_{\R^3}\rho_{\Gamma^n}\phi_j\right)=\str_{P^0}(D^0\Gamma)+\alpha 
\int_{\R^3}\rho_{\Gamma}\phi_j,$$
where we have used that
$$\str_{P^0}(D^0\Gamma^n)=\tr(D^0\Gamma^n_{++})+\tr(D^0\Gamma^n_{--})
=\tr(D^0\Gamma_{++})+\tr(D^0\Gamma_{--})=\str_{P^0}(D^0\Gamma).$$
Let us now pass to the limit in the right hand side. Indeed, we can 
write, by Lemma \ref{trace_blocks},
$$\str_{P^0}(D_j\Gamma^n)=\str_{P'_j}(D_j\Gamma^n)=\tr(D_jP'_j\Gamma^nP'_j)+\tr(D_j(1-P'_j)\Gamma^n(1-P'_j))$$
where $P'_j=\chi_{(-\ii;0)}(D_j)$ and since $P'_j-P^0\in\S_2$ by 
\cite{Scharf1}. Now, using \eqref{chgt_proj}, it is easily seen that 
$P'_j\Gamma^n P'_j\to P'_j\Gamma P'_j$ and 
$(1-P'_j)\Gamma^n(1-P'_j)\to (1-P'_j)\Gamma(1-P'_j)$ in $\S_1$ as 
$n\to\ii$, since this terms can be expanded as a sum of trace class 
operators and products of at least two Hilbert-Schmidt operators 
converging strongly in $\S_2$. Since $D_j$ is bounded by the proof of 
Lemma \ref{DQ_bounded}, we obtain that 
$\str_{P'_j}(D_j\Gamma^n)\to_{n\to\ii} 
\str_{P'_j}(D_j\Gamma)=\str_{P^0}(D_j\Gamma)$ by Lemma 
\ref{trace_blocks}.

As a conclusion, we have proved \eqref{egalite2} for all 
$\Gamma\in\mathcal{G}_\Lambda$. To finish the proof, it remains to 
pass to the limit as $j\to+\ii$. Since
$$\int_{\R^3}\rho_{\Gamma}\phi_j=D(\rho_{\Gamma},\rho_j)$$
and $\rho_\Gamma\in\EuFrak{C}$ (recall that 
$\Gamma\in\mathcal{G}_\Lambda$), $\rho_j\to\rho'_Q$ strongly in 
$\EuFrak{C}$ as $j\to\ii$, we may pass to the limit in the left hand 
side of \eqref{egalite2}. To pass to the limit in the right hand 
side, we use again the fact that
$$\str_{P^0}(D_j\Gamma)=\str_{P'_j}(D_j\Gamma)=\tr(D_jP'_j\Gamma 
P'_j)+\tr(D_j(1-P'_j)\Gamma(1-P'_j)).$$
By the results of Klaus-Scharf \cite{Scharf1} (see also the proof of 
Theorem \ref{existence}), it is known that $P'_j- P'\to0$ in $\S_2$, 
since $\rho_j\to\rho'_Q$ in $\EuFrak{C}$. Using again 
\eqref{chgt_proj}, it is then easily seen that $P'_j\Gamma P'_j\to 
P'\Gamma P'$ and $(1-P'_j)\Gamma(1-P'_j)\to(1-P')\Gamma(1-P')$ in 
$\S_1$ as $j\to\ii$. Since $D_j\to D$ in $\S_\ii$ by Lemma 
\ref{DQ_bounded}, we may thus pass to the limit and obtain the 
desired equality \eqref{egalite}. \qed

\subsection{End of the proof of Theorem \ref{stability}}
We start by proving $1)\Rightarrow 2)$. We thus consider a projector 
$P$ that satisfies the assumption of the Theorem, and is also a 
solution to the equation $P=\chi_{(-\ii;0)}(D_Q)$. We fix some 
$\Gamma\in\mathcal{G}_\Lambda$ and show that 
$\mathcal{E}(\Gamma)\geq\mathcal{E}(Q)$. To this end, we write 
$\mathcal{E}(\Gamma)=\mathcal{E}(Q+\Gamma')$ where 
$\Gamma'=\Gamma-Q=\Gamma+P^0-P$. By assumption, $\Gamma$ fulfills 
$-P^0\leq\Gamma\leq1-P^0$, and so $\Gamma'$ fulfills 
$-P\leq\Gamma'\leq 1-P$. Using Lemma \ref{lemme_egalite}, we may 
expand $\mathcal{E}(Q+\Gamma')$ and obtain
$$
\mathcal{E}(Q+\Gamma') = \str_{P^0}\left(D_Q\Gamma'\right) 
+\frac{\alpha}{2}D(\rho_{\Gamma'},\rho_{\Gamma'})-\frac{\alpha}{2}\int\!\!\!\int\frac{|\Gamma'(x,y)|^2}{|x-y|}dx\,dy+\mathcal{E}(Q).
$$
Using now Lemma \ref{trace_blocks}, we see that it is thus sufficient to prove that
$$\str_{P}\left(D_Q\Gamma'\right) 
+\frac{\alpha}{2}D(\rho_{\Gamma'},\rho_{\Gamma'})-\frac{\alpha}{2}\int\!\!\!\int\frac{|\Gamma'(x,y)|^2}{|x-y|}dx\,dy> 
0,$$
for any $\Gamma'\in\S_1^{P}(\mathcal{H}_\Lambda)$ such that $\Gamma'\neq0$, $\rho_{\Gamma'}\in\EuFrak{C}$ and $-P\leq\Gamma'\leq1-P$,
which is an easy adaptation of the proof of \cite[Theorem 2]{HF1}.

\medskip

We now show $2)\Rightarrow 1)$. Let be $P$ which satisfies the 
assumption of the Theorem, and such that $Q=P-P^0$ is a minimizer of 
$\mathcal{E}$ in $\mathcal{B}_{\Lambda}$. We therefore have, by formula \eqref{separation},
\begin{equation}
\label{energie_gamma}
\tr\left(D_Q\gamma\right) 
+\frac{\alpha}{2}D(\rho_{\gamma},\rho_{\gamma})-\frac{\alpha}{2}\int\!\!\!\int\frac{|\gamma(x,y)|^2}{|x-y|}dx\,dy\geq 
0
\end{equation}
for all $\gamma\in\S_1(\mathcal{H}_\Lambda)$ such that $-P\leq\gamma\leq 1-P$. The proof of Theorem 4 by Bach {\it et al.} \cite{HF1} now implies that $P=\chi_{(-\ii;0)}(D_Q)$.
Their proof is done with $D^{\alpha\phi}$ instead of $D_Q$ but they 
also mention that it can be extended to a more general case, provided 
$0\notin\sigma(D_Q)$ and $P$, $1-P$ leave the domain of $D_Q$ 
invariant, which is the case here. \qed


\section{Proof of Theorem \ref{existence}}
\label{proofs2}
In this section, we prove Theorem \ref{existence} by using a Banach 
fixed-point method.

\subsection{Preliminaries}\label{prelim}
We start by defining the norms and spaces that will be used to apply 
this well known result. In fact, one of the main difficulties we faced in this work consisted in finding suitable Banach spaces.

\subsubsection{Norms and spaces}
We choose the following norms
$$\normQ{Q}:= \left(\int\!\!\!\int 
E(p-q)^2E(p+q)|\widehat{Q}(p,q)|^2\,dp\,dq\right)^{1/2},$$
$$\normR{R}:= \left(\int\!\!\!\int 
\frac{E(p-q)^2}{E(p+q)}|\widehat{R}(p,q)|^2\,dp\,dq\right)^{1/2},$$
$$\normr{\rho}:=\left(\int 
\frac{E(k)^{2}}{|k|^2}|\widehat{\rho}(k)|^2\,dk\right)^{1/2},$$
$$\normp{\phi}:=\left(\int 
|k|^2E(k)^{2}|\widehat{\phi}(k)|^2\,dk\right)^{1/2},$$
where
$$E(k)=\sqrt{1+|k|^2}$$
and denote by $\mathcal{Q}$, $\mathcal{R}$, ${\EuFrak{C}}$ and 
${\mathcal{Y}}$ the associated Hilbert spaces.
The dual space $\EuFrak{C}'$ of $\EuFrak{C}$ will be also useful 
and we introduce
$$\normd{\zeta}:=\left(\int \frac{|k|^2}{E(k)^{2}}|\hat{\zeta}(k)|^2\,dk\right)^{1/2}.$$

In the following, it will be easier to use the norm $\normR{R}$ where 
$R=\frac{Q(x,y)}{|x-y|}$ in our estimates and a relation with 
$\normQ{Q}$ will then be needed. To this end, we first need the 
following well known Lemma, which will be useful throughout the rest 
of the proof.

\begin{lemma}
\label{peetre}
For all $\xi$ and $\eta$ in $\R^3$, we have
\begin{equation}
\label{eq:peetre_sum}
\forall s\geq 0,\ E(\xi)^s\leq 2^{\delta(s)}\left( 
E(\xi-\eta)^s+E(\eta)^s\right)
\end{equation}
\begin{equation}
\label{eq:peetre_prod}
\forall s\in \R,\ E(\xi)^s\leq 2^{|s|}E(\xi-\eta)^sE(\eta)^{|s|},
\end{equation}
with
$$\delta(s)=\left\{\begin{array}{lll}
s & \mbox{ if } & 0\leq s<1\\
s-1 & \mbox{ if } & s\geq 1
\end{array}. \right.$$
\end{lemma}

\begin{remark}\rm
A trivial consequence of 
$(\ref{eq:peetre_sum})$ is the following inequality
\begin{equation}
\label{eq:estim_sum}
\frac{1}{E(p)+E(q)}\leq \min\left(\frac{1}{E(p+q)},\frac{1}{E(p-q)}\right).
\end{equation}
\end{remark} 

\bigskip

We shall also need the following 
\begin{lemma}
We have
\begin{equation}
\sup_{p,q\in\R^3}\frac{E(p+q)}{E(p)^2E(p-q)^2}\leq 2.
\label{estim_fraction}
\end{equation} 
\end{lemma} 

\noindent{\it Proof.} Let us introduce the function $f(p,q)=E(p+q)E(p)^{-2}E(p-q)^{-2}$ for $(p,q)\in\R^3\times\R^3$. We have
$$0\leq f(p,q)=\frac{E(2p-(p-q))}{E(p)^2E(p-q)^2}\leq \frac{2E(2p)E(p-q)}{E(p)^2E(p-q)^2}\leq \frac{4}{E(p)E(p-q)}$$
by Lemma \ref{peetre}.
Therefore $\lim_{(p,q)\to\ii}f(p,q)=0$ and $f$ attains its maximum on $\R^3\times\R^3$. Computing $\nabla_qf(p,q)$, we see that at a critical point of $f$, $p$ and $q$ are always parallel. It therefore suffices to study the function $g(x,y)=E(x+y)E(x)^{-2}E(x-y)^{-2}$ for $(x,y)\in\R\times\R$. It is then easy to see that $\max_{\R^2}g < 2$ (the bound \eqref{estim_fraction} is indeed not optimal). \qed

\medskip

Now we can give a connection between $\normR{R}$ and $\normQ{Q}$ when 
$R=\frac{Q(x,y)}{|x-y|}$ (we also recall the easy relation between 
$\normr{\rho}$ and $\normp{\phi}$ when 
$\phi=\rho\ast\frac{1}{|\cdot|}$).
\begin{lemma}
\label{estim_RQ}
If $\rho\in\EuFrak{C}$ and $Q\in\mathcal{Q}$, then we have 
$\phi=\rho\ast\frac{1}{|\cdot|}\in\mathcal{Y}$ and 
$R(x,y)=\frac{Q(x,y)}{|x-y|}\in\mathcal{R}$ and more precisely
$$\normp{\phi}=4\pi\normr{\rho},$$
\begin{equation}
\label{eq:estim_RQ}
\normR{R}\leq C_R\normQ{Q},
\end{equation}
\begin{equation}
\norm{R\,|D_0|^{-1}}_{\S_2}\leq \sqrt{2}\normR{R},
\label{estim_R_D0}
\end{equation} 
with
\begin{equation}
\label{CR}
C_R:= 
\frac{1}{2\pi^2}\inf_{\theta\in(0;2)}\sup_{x\in\R^3}\left(E(2x)^{\theta}\int_{\R^3}\frac{du}{E(2u)^{1+\theta}|u-x|^2}\right).
\end{equation}
\end{lemma}

\noindent {\it Proof.} We have
$$\widehat{R}(p,q)=\frac{1}{2\pi^2}\int_{\R^3}\frac{\widehat{Q}(p-l,q-l)}{|l|^2}\,dl$$
so we obtain, for some fixed $\theta\in(0;2)$
\begin{eqnarray*}
\normR{R}^2 & = & \int\!\!\!\int 
\frac{E(p-q)^2}{E(p+q)}|\widehat{R}(p,q)|^2\,dp\,dq\\
  & = & 8\int\!\!\!\int \frac{E(2v)^2}{E(2u)}|\widehat{R}(u+v,u-v)|^2\,du\,dv\\
  & = & \frac{8}{(2\pi^2)^2}\int\!\!\!\int\!\!\!\int\!\!\!\int 
\frac{E(2v)^2}{E(2u)}\frac{\widehat{Q}(l+v,l-v)\cdot\widehat{Q}(l'+v,l'-v)}{|l-u|^2|l'-u|^2}\,du\,dv\,dl\,dl'\\
  & = & \frac{8}{(2\pi^2)^2}\int\!\!\!\int\,du\,dv 
\frac{E(2v)^2}{E(2u)}\times\\
 & & \qquad\times\int\!\!\!\int\frac{E(2l)^{\frac{1+\theta}{2}}\widehat{Q}(l+v,l-v)\,E(2l')^{\frac{1+\theta}{2}}\widehat{Q}(l'+v,l'-v)}{E(2l')^{\frac{1+\theta}{2}}|l-u|\,|l'-u|\,E(2l)^{\frac{1+\theta}{2}}|l-u|\, 
|l'-u|}dl\,dl'\\
  & \leq & \frac{8}{(2\pi^2)^2}\int\!\!\!\int\!\!\!\int\!\!\!\int 
\frac{E(2v)^2}{E(2u)}\frac{E(2l)^{1+\theta}}{E(2l')^{1+\theta}}\frac{|\widehat{Q}(l+v,l-v)|^2}{|l-u|^2|l'-u|^2}\,du\,dv\,dl\,dl'\\
  & \leq & 8\int\!\!\!\int E(2v)^2E(2l) |\widehat{Q}(l+v,l-v)|^2 
K_\theta(l) \,dv\,dl
\end{eqnarray*}
where
$$ K_\theta(l):= \frac{E(2l)^{\theta}}{(2\pi^2)^2}\int\!\!\!\int 
\frac{1}{E(2u) E(2l')^{1+\theta}|l-u|^2|l'-u|^2}\,du\,dl'.$$
Now, let us introduce
$$C_\theta:=\sup_{x\in\R^3}\left(E(2x)^{\theta}\int_{\R^3}\frac{du}{E(2u)^{1+\theta}|u-x|^2}\right).$$
Remark that
$$\int_{\R^3}\frac{du}{E(2u)^{1+\theta}|u-x|^2}\leq 
\frac{1}{2^{1+\theta}|x|^\theta}\int_{\R^3}\frac{du}{|u|^{1+\theta}|u-e_x|^2}$$
where $e_x:=x/|x|$, showing that $C_\theta<\ii$ when 
$\theta\in(0;2)$. Now we have
\begin{eqnarray*}
  K_\theta(l) &=& \frac{E(2l)^{\theta}}{(2\pi^2)^2}\int\,du 
\frac{1}{E(2u)^{1+\theta} 
|l-u|^2}\left(E(2u)^\theta\int\,\frac{1}{E(2l')^{1+\theta}|l'-u|^2}dl'\right)\\
  & \leq & \frac{E(2l)^{\theta}}{(2\pi^2)^2}\int\,du 
\frac{1}{E(2u)^{1+\theta} |l-u|^2}\times C_\theta  \leq 
\left(\frac{C_\theta}{2\pi^2}\right)^2
\end{eqnarray*}
and so
$$\normR{R}^2\leq 
8\left(\frac{C_\theta}{2\pi^2}\right)^2\int\!\!\!\int E(2v)^2E(2l) 
|\widehat{Q}(l+v,l-v)|^2 \,dv\,dl\leq 
\left(\frac{C_\theta}{2\pi^2}\right)^2\normQ{Q}^2$$
which ends the proof of $(\ref{eq:estim_RQ})$.

To prove \eqref{estim_R_D0}, we remark that we have, by \eqref{estim_fraction},
$$
\norm{R\;|D_0|^{-1}}_{\S_2}^2  = \iint\frac{|\widehat{R}(p,q)|^2}{E(p)^2}\,dp\,dq \leq 2\iint
\frac{E(p-q)^2|\widehat{R}(p,q)|^2}{E(p+q)}\,dp\,dq =2\normR{R}^2. \qed
$$

\subsubsection{An estimate from below for $D_Q$}
We now state a Lemma in which we give a lower estimate for the operator
$$D_{Q,\mu}:=D^{0}+\alpha\mu\ast\frac{1}{|r|}-\alpha\frac{Q(x,y)}{|x-y|}$$
by $D^0$, in terms of the spaces introduced above.

For our result, we are interested in $D_Q=D_{Q,\rho_Q-n}$ but this 
definition with an arbitrary density $\mu$ will be useful later on.
\begin{lemma}
\label{estim_DQ}
Assume that $(Q,\mu)\in\mathcal{Q}\times\EuFrak{C}$ are such that
\begin{equation}
\alpha\left(2\sqrt{\pi}\normr{\mu}+\sqrt{2}C_R\normQ{Q}\right)<1.
\label{cond_DQ}
\end{equation}
Then $D_{Q,\mu}$ is a bounded operator which satisfies
\begin{equation}
|D_{Q,\mu}|\geq 
\left(1-\alpha\left(2\sqrt{\pi}\normr{\mu}+\sqrt{2}C_R\normQ{Q}\right)\right)|D^0|.
\label{estim_DQD0}
\end{equation}
\end{lemma}

\noindent {\it Proof.} We have, with $\phi'=\mu\ast\frac{1}{|\cdot|}$,
$$\norm{\phi' u}_{L^2}\leq \norm{\phi'}_{L^\ii}\norm{u}_{L^2}\leq 
2\pi^{1/2}\normr{\mu}\norm{|D_0|\cdot u}_{L^2}$$
by Lemma \ref{injections}, and
$$\norm{Ru}_{L^2}=\norm{R|D_0|^{-1}|D_0|u}_{L^2}\leq 
\norm{R|D_0|^{-1}}_{\S_2}\norm{|D_0| u}_{L^2}\leq 
\sqrt{2}C_R\normQ{Q}\norm{|D_0|\cdot u}_{L^2}$$
by \eqref{estim_R_D0}. This shows that $|\phi'-R|\leq 
(2\pi^{1/2}\normr{\mu}+\sqrt{2}C_R\normQ{Q})|D^0|$, the square root 
being monotone. This proves that $D_{Q,\mu}$ is bounded since $D_0$ 
is bounded on $\mathcal{H}_\Lambda$, and gives the expected 
inequality. \qed

\medskip

Remark that Lemma \ref{estim_DQ} will be useful when we shall apply 
Theorem \ref{stability} (see the condition \eqref{estim_d} in the 
statement).  It also implies $0\notin\sigma(D_{Q,\mu})$, a fact that 
will be used to compute the projection $\chi_{(-\ii;0)}(D_{Q,\mu})$.

\subsubsection{Expansion by Cauchy's formula}\label{Cauchy_formula}
We want to solve the equation
$$Q=\chi_{(-\ii;0)}\left(D^0+\alpha(\rho_Q-n)\ast\frac{1}{|\cdot|}-\alpha 
\frac{Q(x,y)}{|x-y|}\right)-\chi_{(-\ii;0)}(D^0):=F_1(Q).$$

If $\alpha\left(2\sqrt{\pi}\normr{\rho_Q-n}+\sqrt{2}C_R\normQ{Q}\right)<1$, 
then $0\notin\sigma(D_Q)$ by Lemma \ref{estim_DQ}. We may thus use 
the method of \cite{HS} and expand $F_1$ by Cauchy's formula
$$F_1(Q)=-\frac{1}{2\pi}\int_{-\ii}^{+\ii}d\eta\left( 
\frac{1}{D_Q+i\eta}-\frac{1}{D^0+i\eta}\right)=\sum_{n=1}^\ii 
\alpha^nQ_n$$
where
$$Q_n=-\frac{1}{2\pi}\int_{-\ii}^{+\ii}d\eta\frac{1}{D^0+i\eta}\left((R_Q-\phi'_Q)\frac{1}{D^0+i\eta}\right)^n$$
$$\rho'_Q=\rho_Q-n,\ \ \phi'_Q=\rho'_Q\ast\frac{1}{|r|},\ \ 
R_Q(x,y)=\frac{Q(x,y)}{|x-y|}.$$
We shall write
$$Q_n=\sum_{k,l/\ k+l=n} Q_{k,l}$$
$$Q_{k,l}=\frac{(-1)^{l+1}}{2\pi} \sum_{\substack{I\cup J =\{1,...,n\},\\ 
|I|=k,\ |J|=l}} 
\int_{-\ii}^{+\ii}d\eta\frac{1}{D^0+i\eta}\prod_{j=1}^n 
\left(R_j\frac{1}{D^0+i\eta}\right)$$
where $R_j=R_Q$ if $j\in I$ and $R_j=\phi'_Q$ if $j\in J$ ($Q_{k,l}$ 
is the sum of all the terms containing $k$ $R_Q$'s and $l$ 
$\phi'_Q$'s). We also denote $\rho_{k,l}:=\rho_{Q_{k,l}}$.

Hence our equation can be written
\begin{equation}
\label{equation_Qrho}
\left\{\begin{matrix}
\displaystyle Q=\sum_{n=1}^\ii \alpha^nQ_n(Q,\rho'_Q)\\
\displaystyle \rho_Q=\sum_{n=1}^\ii \alpha^n\rho_n(Q,\rho'_Q),\\
\end{matrix} \right.
\end{equation}
where we recall that $Q_n$ and $\rho_n$ depend on both $Q$ and $\rho'_Q=\rho_Q-n$.
In order to have a better condition on $\alpha$ and $\Lambda$, we 
shall now change the second equation for the density, by taking into 
account the special form of the first order term $\rho_1$. To this 
end, we need to compute this term explicitely.

\subsubsection{The first order density}\label{first_order_density}
Recall that
$$Q_{0,1}=\frac{1}{2\pi}\int_{-\ii}^{+\ii}d\eta 
\frac{1}{D^0+i\eta}\phi'_Q\frac{1}{D^0+i\eta}$$
so that
$$\widehat{Q_{0,1}}(p,q)=(2\pi)^{-5/2} \int_{-\ii}^{+\ii}d\eta 
\frac{1}{\alp\cdot 
p+\beta+i\eta}\widehat{\phi'_Q}(p-q)\frac{1}{\alp\cdot 
q+\beta+i\eta}.$$ We now introduce
\begin{eqnarray}
M(p,q) & := & \frac{1}{\pi}\int_{-\ii}^{+\ii}d\eta \frac{1}{\alp\cdot 
p+\beta+i\eta}\cdot\frac{1}{\alp\cdot q+\beta+i\eta}\nonumber\\
  & = & \frac{1}{\pi}\int_{-\ii}^{+\ii}d\eta \frac{\alp\cdot 
p+\beta-i\eta}{p^2+1+\eta^2}\cdot \frac{\alp\cdot 
q+\beta-i\eta}{q^2+1+\eta^2}\nonumber\\
  & = & \frac{1}{E(p)+E(q)}\left(\frac{(\alp\cdot 
p+\beta)}{E(p)}\frac{(\alp\cdot 
q+\beta)}{E(q)}-1\right).\label{def_matrix_M}
\end{eqnarray}
Hence
$$\widehat{Q_{0,1}}(p,q)=\frac{1}{2^{5/2}\pi^{3/2}}\widehat{\phi'_Q}(p-q) 
M(p,q).$$

This enables us to compute
\begin{eqnarray}
\widehat{\rho_{0,1}}(k) & = & 
\frac{1}{(2\pi)^{3/2}}\int_{|l|\leq\Lambda} 
\Tr{\widehat{Q_{0,1}}(l+k/2,l-k/2)}dl\nonumber\\
& = & \frac{1}{16\pi^3}\widehat{\phi'_Q}(k)\int_{|l|\leq\Lambda} 
\Tr{M(l+k/2,l-k/2)}dl\nonumber\\
  &= & -\frac{1}{4\pi}\widehat{\phi'_Q}(k)|k|^2B_\Lambda(k) \nonumber\\
& = & -\widehat{\rho'_Q}(k)B_\Lambda(k)\label{Order1_B}
\end{eqnarray}
where
\begin{equation}
B_\Lambda(k)=-\frac{1}{\pi^2 |k|^2}\int_{|l|\leq\Lambda} 
\frac{(l+k/2)\cdot(l-k/2)+1-E(l+k/2)E(l-k/2)}{E(l+k/2)E(l-k/2)(E(l+k/2)+E(l-k/2))}dl.
\label{def_B_Lambda}
\end{equation} 

This function is computed in \cite{PauliRose}
$$B_\Lambda(k)=\frac{1}{\pi}\int_0^{\frac{\Lambda}{E(\Lambda)}}\frac{z^2-z^4/3}{1-z^2}\frac{dz}{1+|k|^2(1-z^2)/4}$$
and it is logarithmically divergent since
$$B_\Lambda(0)=\frac{1}{\pi}\int_0^{\frac{\Lambda}{E(\Lambda)}}\frac{z^2-z^4/3}{1-z^2}\,dz=\frac{2}{3\pi}\log(\Lambda)-\frac{5}{9\pi}+\frac{2}{3\pi}\log 
2 + O(1/\Lambda^2).$$

\subsubsection{Equation}
We are now able to introduce the function on which we shall apply the 
fixed-point theorem. According to what we said above, the equation in 
$\rho_Q$ can be written
\begin{equation}
\label{equ_rho_1}
\rho_Q=\sum_{n=1}^\ii \alpha^n\rho_n(Q,\rho'_Q)
\end{equation}
or equivalently (we forget the dependence in $Q$ and $\rho'_Q$ for simplicity)
$$\widehat{\rho_Q}(k)=-\alpha 
B_\Lambda(k)\widehat{\rho'_Q}(k)+\alpha\widehat{\rho_{1,0}}(k)+\sum_{n=2}^\ii 
\alpha^n\widehat{\rho_n}(k)$$
and
\begin{equation}
\label{equ_rho_2}
\widehat{\rho'_Q}(k)=-\frac{1}{1+\alpha 
B_\Lambda(k)}\widehat{n}(k)+\frac{1}{1+\alpha 
B_\Lambda(k)}\left(\alpha\widehat{\rho_{1,0}}(k)+\sum_{n=2}^\ii 
\alpha^n\widehat{\rho_n}(k)\right),
\end{equation}
which is more adapted to a fixed-point argument, since the divergent term $B_\Lambda(k)$ appears now in the denominator.

Notice that we now study an equation for the full density $\rho'_Q=\rho_Q-n$ and not for $\rho_Q$ as previously. We therefore introduce the following space:
$$\mathcal{X}=\mathcal{Q}\times\EuFrak{C}$$
consisting of all the pairs $(Q,\rho')$ such that $Q\in \mathcal{Q}$ 
and $\rho'\in\EuFrak{C}$. Notice that in this space, $\rho'$ can be 
different from $\rho_Q-n$. However, we shall find a solution of the 
equations in this space, which satisfies $\rho'=\rho_Q-n$. We also 
introduce on $\mathcal{X}$ the norm
$$\norm{(Q,\rho')}=C_R\sqrt{2}\normQ{Q}+2\sqrt{\pi}\normr{\rho'},$$
where we recall that $C_R$ is defined in Lemma \ref{estim_RQ}. In the following, we shall keep the notation $\rho'$ to remind the reader that the equation indeed concerns $\rho_Q'$ and not $\rho_Q$.

We now introduce the function $F:\mathcal{X}\to\mathcal{X}$ defined by
$$F(Q,\rho')=\bigg(F_Q(Q,\rho')\ ,\ F_\rho(Q,\rho')\bigg)$$
where
\begin{equation}
\label{eq_Q}
F_Q(Q,\rho')=\chi_{(-\ii;0)}(D_{Q,\rho'})-P^0=\sum_{n=1}^\ii \alpha^nQ_n(Q,\rho')
\end{equation}
\begin{equation}
\label{eq_rho}
\widehat{F_\rho(Q,\rho')}=-\frac{1}{1+\alpha 
B_\Lambda}\widehat{n}+\frac{1}{1+\alpha 
B_\Lambda}\left(\alpha\widehat{\rho_{1,0}}(Q,\rho')+\sum_{n=2}^\ii 
\alpha^n\widehat{\rho_n}(Q,\rho')\right),
\end{equation}
$Q_n(Q,\rho')$ and $\rho_n(Q,\rho')$ being defined in 
\eqref{equation_Qrho} (replace $\rho'_Q$ by $\rho'$). Remark that 
$\rho_n=\rho_{Q_n}$ for all $n\geq 2$. In the proof of Theorem 
\ref{existence}, we solve the fixed-point equation in $\mathcal{X}$
$$F(Q,\rho')=(Q,\rho').$$

\subsection{Existence of a fixed-point of $F$}
To prove our main Theorem, we need the following estimates
\begin{prop}
\label{estimates}
Assume that $(Q,\rho')\in\mathcal{X}$ is such that 
$0\notin\sigma(D_{Q,\rho'})$. Then we have
\begin{equation}
\norm{F(Q,\rho')}\leq 
2\sqrt{\pi} \normr{n}+\kappa_1(\Lambda)\alpha\norm{(Q,\rho')} + 
\sum_{n=2}^{+\ii}\kappa_n\left(\alpha\norm{(Q,\rho')}\right)^n
\label{estim_F}
\end{equation}
\begin{equation}
\norm{F'(Q,\rho')}\leq \kappa_1(\Lambda)\alpha + 
\alpha\sum_{n=2}^{+\ii}n\kappa_n\left(\alpha\norm{(Q,\rho')}\right)^{n-1}
\label{estim_dF}
\end{equation}
where
$$\kappa_1(\Lambda)=\max\left( 
\frac{C_R\sqrt{2}}{\sqrt{\pi}}\sqrt{\log\Lambda}\, , \, 
\sqrt{2}C_R+\frac{\sqrt{\log\Lambda}}{2^{3/2}\sqrt{\pi}}\right)\sim_{\Lambda\to\ii}\frac{C_R\sqrt{2}}{\sqrt{\pi}}\sqrt{\log\Lambda}$$
and $(\kappa_n)_{n\geq 2}$ is a sequence of positive numbers 
independent of $\Lambda$ and which satisfies 
$\kappa_n\sim_{n\to\ii}K\sqrt{n}$ for some constant $K$.
\end{prop}

To prove this proposition, we have to do some tedious estimates.
Before starting this proof, let us show that Theorem \ref{existence} follows from Proposition \ref{estimates}. 

\medskip

\noindent{\it Proof of Theorem \ref{existence}.} 
We introduce the function $f(x)=\sum_{n=2}^\ii\kappa_nx^n$, which 
is a power series with a radius of convergence equal to 1. The 
estimates $(\ref{estim_F})$ and $(\ref{estim_dF})$ can be written
\begin{equation}
\label{ineg_prem}
\norm{F(Q,\rho')}\leq 
2\sqrt{\pi} \normr{n}+\kappa_1(\Lambda)\alpha \norm{(Q,\rho')}+
 f\left(\alpha\norm{(Q,\rho')}\right)
\end{equation}
$$\norm{F'(Q,\rho')}\leq \kappa_1(\Lambda)\alpha + \alpha f'\left(\alpha\norm{(Q,\rho')}\right).$$
To apply the Banach fixed-point theorem, we now have to find a ball 
$B(0,R)\subset\mathcal{X}$ which is invariant under the function $F$ 
and on which $F$ is a contraction. Let $R>0$ be some fixed radius. We 
have
$$\sup_{(Q,\rho')\in B(0,R)}\norm{F'(Q,\rho')}\leq 
\kappa_1(\Lambda)\alpha + \alpha f'(\alpha R):=\mu.$$
Moreover, we also have
$$\norm{F(Q,\rho')}\leq \norm{F(Q,\rho')-F(0,0)}+\norm{F(0,0)}\leq 
\mu\norm{(Q,\rho')}+\norm{F(0,0)}.$$
Therefore a condition for the ball $B(0,R)$ to be invariant under the 
action of $F$ is $\norm{F(0,0)}\leq (1-\mu)R$. Notice that since $F(0,0)\neq(0,0)$, this inequality also contains the contraction condition $\mu<1$. Additionally due to 
Lemma \ref{estim_DQ} we assume $\alpha R<1$ as well as
$$\frac{\alpha\pi}{4(1-\alpha R)}\leq 1,$$ due to Theorem 
\ref{stability} and Lemma \ref{estim_DQ}
which is equivalent to
$$\alpha\leq \frac{1}{\pi/4+R}.$$

As a conclusion, if $(\alpha,R)$ fulfills
\begin{equation}\left\{ \begin{array}{l}\displaystyle
2\sqrt{\pi} \normr{n}+ \alpha R \kappa_1(\Lambda) + \alpha R 
f'(\alpha R) \leq R\\
\displaystyle \alpha\leq \frac{1}{\pi/4+R},
\end{array}\right.
\label{ineq_sol0}
\end{equation}
then we are able to apply the Banach Fixed-Point Theorem on $B(0,R)$. 
Remark that these inequalities also contain the conditions $\mu<1$ 
and $\alpha R<1$. Notice also that if $(\alpha, R)$ is a 
solution to \eqref{ineq_sol0}, then $(\alpha',R)$ is a solution to 
\eqref{ineq_sol0} for all $\alpha'\leq\alpha$, since the function 
which appears on the left of \eqref{ineq_sol0} is increasing in 
$\alpha$.

Now, if we assume that $2\sqrt{\pi}\alpha \normr{n}\leq b$, we obtain 
that if $(\alpha,R)$ fulfills
\begin{equation}\left\{ \begin{array}{l}\displaystyle
\frac{b}{\alpha}+ \alpha R \kappa_1(\Lambda) + \alpha R 
f'(\alpha R) \leq R\\
\displaystyle \alpha\leq \frac{1}{\pi/4+R},
\end{array}\right.
\label{ineq_sol}
\end{equation}
then it also fulfills \eqref{ineq_sol0}. 
The first inequation of \eqref{ineq_sol} is simpler when it is 
written in terms of the variables $\alpha$ and $x:=\alpha R$. It 
becomes
\begin{equation}
\label{ineq_sol2}
\frac{b}{\alpha}+ \kappa_1(\Lambda)x + x f'(x) \leq \frac{x}{\alpha}
\end{equation}
which implies $x\in[b;1)$. 
Now, given $b$, $\Lambda$ and $x$ let us 
call $a_{b,\Lambda}(x)$ the maximal value of $\alpha$ such that \eqref{ineq_sol2} holds, i.e.
\begin{equation}
\label{fn_a}
a_{b,\Lambda}(x)=\frac{x-b}{\kappa_1(\Lambda)x + x f'(x)},
\end{equation}
which is defined for $x$ in $[b;1)$. Since $\lim_{x\to1}a_{b,\Lambda}(x)=a_{b,\Lambda}(b)=0$, we may denote by $x_\Lambda\in(b;1)$ the largest maximizer of the function $a_{b,\Lambda}$ in the interval $[b;1)$.

We now define $R_{b}(\Lambda):=x_\Lambda / a_{b,\Lambda}(x_\Lambda)$
and
$$\alpha_b(\Lambda):=\min\left(a_{b,\Lambda}(x_\Lambda),\frac{1}{\pi/4+R_{b}(\Lambda)}\right).$$

As a conclusion, for all $0\leq\alpha\leq \alpha_b(\Lambda)$, $(\alpha,R_{b}(\Lambda))$ is a 
solution of \eqref{ineq_sol0}. This means that $F$ is a contraction 
on $B(0,R_{b}(\Lambda))$, on which we can apply Banach Theorem. This gives a 
unique solution to the equation $F(Q,\rho')=(Q,\rho')$ in 
$B(0,R_{b}(\Lambda))\subset\mathcal{X}$. 

Let us now show that $P$ is indeed a solution to \eqref{equation_P}. 
In fact $\rho'$ is a solution to \eqref{equ_rho_2} and so $\rho=\rho'+n$ is a 
solution to \eqref{equ_rho_1}. On the other hand, we have 
$Q=\chi_{(-\ii;0)}(D_{Q,\rho'})-P^0$, and 
\eqref{equ_rho_1} means exactly that $\rho=\rho_Q$. Hence, $P$ is a 
solution to $P=\chi_{(-\ii;0)}(D_Q)$. Thanks to the proof, we know 
that $P$ satisfies the assumptions of Theorem \ref{stability}, and so 
$P$ is the unique BDF-stable vacuum (i.e. $P-P^0$ is the unique global minimizer of the BDF energy).

\medskip

To end the proof, let us study the behaviour of $\alpha_b(\Lambda)$ as $\Lambda\to\ii$. 
Computing ${\rm d}a_{b,\Lambda}(x)/{\rm d}x$, we find that $x_\Lambda$ must satisfy the equation
$$\kappa_1(\Lambda)+f'(x_\Lambda)=\frac{x_\Lambda(x_\Lambda-b)}{b}f''(x_\Lambda).$$
Since $\kappa_1(\Lambda)$ diverges as $\Lambda\to\ii$, we see that $f''(x_\Lambda)\to\ii$ and therefore $x_\Lambda\to1$ as $\Lambda\to\ii$. Now, since $f'(x)=o_{x\to1}\left(f''(x)\right)$, we obtain that $f'(x_\Lambda)=o_{\Lambda\to\ii}(\kappa_1(\Lambda))$. Thus
$$\alpha_{b,\Lambda}(x_\Lambda)\sim_{\Lambda\to\ii}\frac{1-b}{\kappa_1(\Lambda)}\qquad \text{ and }\qquad R_{b}(\Lambda)\sim_{\Lambda\to\ii}\frac{\kappa_1(\Lambda)}{1-b}.$$
As a conclusion,
$$\alpha_b(\Lambda)\sim_{\Lambda\to\ii}\frac{1-b}{\kappa_1(\Lambda)}\sim_{\Lambda\to\ii}\frac{\sqrt{\pi}(1-b)}{C_R\sqrt{2}\sqrt{\log\Lambda}}=\frac{C(1-b)}{\sqrt{\log\Lambda}}$$
with $C=\frac{\sqrt{\pi}}{\sqrt{2}C_R}$.\qed

\subsection{Proof of Proposition \ref{estimates}: estimates}
In this section, we prove the claimed estimates of Proposition \ref{estimates}. We will have to introduce many constants. For the sake of clarity, a guide is provided to the reader at the very end of the proof, section \ref{def_constantes}.

Remark first that we have
$$
\frac{1}{1+\alpha B_\Lambda(k)}\leq 1$$
for all $k\in\R^3$. Therefore, to estimate the norm 
$\normr{F_\rho(Q,\rho')}$, it suffices to estimate the norms of 
$\rho_{1,0}(Q,\rho')$ and $\rho_n(Q,\rho')$, due to \eqref{eq_rho}.

For $(Q,\rho')\in\mathcal{X}$, we introduce the notation $R=\frac{Q(x,y)}{|x-y|}\in\mathcal{R}$, and 
$\phi'=\rho'\ast\frac{1}{|\cdot|}$.
We then remark that
$$\norm{F(Q,\rho')}\leq 2\sqrt{\pi}\normr{n}+\alpha\norm{(Q_1,\rho_{1,0})}+\sum_{n\geq2}\alpha^n\norm{(Q_n,\rho_n)}$$
and estimate each term separately. A similar argument can be done for 
$F'(Q,\rho')$.

\subsubsection{First order terms}
\begin{lemma}
\label{order1}
We have the following estimates:
$$\normQ{Q_{0,1}}\leq 
\frac{(\log\Lambda)^{1/2}}{2\pi}\normp{\phi'}=2(\log\Lambda)^{1/2}\normr{\rho'},$$
$$\normQ{Q_{1,0}} \leq \normR{R_Q}\leq C_R \normQ{Q}, \qquad 
\normr{\rho_{1,0}}\leq \frac{C_R(\log\Lambda)^{1/2}}{4\pi}\normQ{Q}.$$
Therefore
$$\norm{(Q_1,\rho_{1,0})}\leq \kappa_1(\Lambda)\norm{(Q,\rho')},$$
where
\begin{equation}
\kappa_1(\Lambda)=\max\left( 
\frac{C_R\sqrt{2}}{\sqrt{\pi}}\sqrt{\log\Lambda}\, , \, 
\sqrt{2}C_R+\frac{\sqrt{\log\Lambda}}{2^{3/2}\sqrt{\pi}}\right).
\label{Kappa_1}
\end{equation} 
\end{lemma}

\noindent{\it Proof.} Recall that
\begin{equation}
\label{Q01}
\widehat{Q_{0,1}}(p,q)=\frac{1}{2^{5/2}\pi^{3/2}}\widehat{\phi'}(p-q) M(p,q),
\end{equation}
where the matrix $M(p,q)$ is defined in \eqref{def_matrix_M}, and 
whose properties are summarized in the following
\begin{lemma}
\label{estim_M}
Let $\Lambda^+(p)=\frac{\alp\cdot p+\beta +E(p)}{2E(p)}$ and 
$\Lambda^-(p)=\frac{-(\alp\cdot p+\beta) +E(p)}{2E(p)}$ be the 
projections matrices in $\C^4$ onto the eigenspaces of $D^0$ in 
Fourier space. We then have
$$\Tr{M(p,q)}  =  -4\frac{1}{E(p)+E(q)}\Tr{\Lambda^+(p)\Lambda^-(q)}$$
$$|M(p,q)|^2=\Tr{M(p,q)M(p,q)^\ast}  = 
8\frac{1}{(E(p)+E(q))^2}\Tr{\Lambda^+(p)\Lambda^-(q)}$$
$$\Tr{\Lambda^+(p)\Lambda^-(q)}=\Tr{\Lambda^-(p)\Lambda^+(q)}=1-\frac{p\cdot 
q+1}{E(p)E(q)}.$$
Moreover, we have
\begin{equation}
\label{eq:estim_M}
\forall p,q\in\R^3,\ \ \ \Tr{\Lambda^+(p)\Lambda^-(q)}\leq 
\min\left(\frac{|p-q|^2}{2E((p+q)/2)^2}\, ,\, 2\right).
\end{equation}
\end{lemma}

\noindent{\it Proof of Lemma \ref{estim_M}.} We only prove 
$(\ref{eq:estim_M})$. We have $\Tr{\Lambda^+(p)\Lambda^-(q)} \leq 
|\Lambda^+(p)|\,|\Lambda^-(q)|=2$, so when 
$t:=\frac{|p-q|^2}{4E((p+q)/2)^2}\geq 1$, there is nothing to 
prove. Now we have
\begin{eqnarray*}
1-\frac{p\cdot q+1}{E(p)E(q)} & = & 1-\frac{l^2-k^2+1}{E(l+k)E(l-k)} 
\mbox{ with } l=\frac{p+q}{2},\ k=\frac{p-q}{2}\\
  & = & 1 - \frac{1-t}{\sqrt{(1+t)^2-4zt}}
\end{eqnarray*}
where $t=|k|^2/E(l)^2$ and $z=\frac{(l\cdot 
k)^2}{|k|^2(1+l^2)}\in[0;1)$. When $t\in[0;1)$ and $z\in[0;1)$, the 
expression above is decreasing in $z$ and so we obtain
$$1-\frac{p\cdot q+1}{E(p)E(q)} \leq 1 - 
\frac{1-t}{\sqrt{(1+t)^2}}=\frac{2t}{1+t}\leq 2t$$
which ends the proof. \qed

\bigskip

\noindent $\bullet$ Let us now treat $Q_{0,1}$. From \eqref{Q01}, we obtain
$$|\widehat{Q_{0,1}}(p,q)|^2= \frac{1}{2^{5}\pi^{3}} 
|\widehat{\phi'}|^2(p-q)|M(p,q)|^2,$$
and so
\begin{multline*}
\int\!\!\!\int E(p-q)^{2}E(p+q)|\widehat{Q_{0,1}}(p,q)|^2dp\,dq\\
  = \frac{1}{2^{5}\pi^{3}}\int\!\!\!\int\,dk\,du 
E(k)^{2}E(2u)|\widehat{\phi'}|^2(k)\chi(|u|\leq\Lambda)|M(u+k/2,u-k/2)|^2\\
  \leq  \frac{4}{2^{5}\pi^{3}}\left(\int dk 
|k|^2E(k)^{2}|\widehat{\phi'}|^2(k)\right)\left(\int_{|u|\leq\Lambda}\frac{1}{E(2u)E(u)^2}\right)
\end{multline*}
by Lemma \ref{estim_M}. Now we have
\begin{equation}
\label{estim_Log}
\int_{|u|\leq\Lambda}\frac{du}{E(2u)E(u)^2}=4\pi 
\left(\frac{1}{2}\mbox{argsh}(2\Lambda)+\frac{1}{\sqrt{3}}\mbox{argth}\left( 
\frac{\sqrt{3}\Lambda}{\sqrt{1+4\sqrt{\Lambda}}}\right)\right) \leq 
2\pi\log\Lambda
\end{equation}
for $\Lambda\geq 3$. So we obtain
$$\normQ{Q_{0,1}}\leq 
\frac{(\log\Lambda)^{1/2}}{2\pi}\normp{\phi'}=2(\log\Lambda)^{1/2}\normr{\rho'}.$$

\bigskip

\noindent $\bullet\ \rho_{1,0}$ and $Q_{1,0}$.
We have
$$Q_{1,0}=-\frac{1}{2\pi}\int_{-\ii}^{+\ii}d\eta 
\frac{1}{D^0+i\eta}R\frac{1}{D^0+i\eta}$$
so that
\begin{eqnarray*}
\widehat{Q_{1,0}}(p,q) & = & -(2\pi)^{-1} \int_{-\ii}^{+\ii}d\eta 
\frac{1}{\alp\cdot p+\beta+i\eta}\widehat{R}(p,q)\frac{1}{\alp\cdot 
q+\beta+i\eta}\\
  & = & -\frac{1}{2}\frac{1}{E(p)+E(q)}\left(\frac{(\alp\cdot 
p+\beta)}{E(p)}\widehat{R}(p,q)\frac{(\alp\cdot 
q+\beta)}{E(q)}-\widehat{R}(p,q)\right)
\end{eqnarray*}
and
$$|\widehat{Q_{1,0}}(p,q)|^2\leq 
\frac{1}{(E(p)+E(q))^2}|\widehat{R}(p,q)|^2\leq 
\frac{1}{E(p+q)^2}|\widehat{R}(p,q)|^2,$$
showing that
$$\normQ{Q_{1,0}} \leq  \normR{R}\leq C_R \normQ{Q}.$$
Now, we have
\begin{eqnarray*}
\widehat{\rho_{1,0}}(k) & = & \frac{1}{(2\pi)^{3/2}}\int_{\R^3} 
\Tr{\widehat{Q_{1,0}}\left(l+\frac{k}{2},l-\frac{k}{2}\right)}dl\\
  & = & -\frac{1}{2^{5/2}\pi^{3/2}}\int_{\R^3} 
\Tr{\widehat{R}\left(l+\frac{k}{2},l-\frac{k}{2}\right) 
M\left(l+\frac{k}{2},l-\frac{k}{2}\right)}\chi(|l|\leq\Lambda)dl,
\end{eqnarray*}
so we obtain
\begin{multline*}
 |\widehat{\rho_{1,0}}(k)| \leq 
\frac{1}{2^{5/2}\pi^{3/2}}\left(\int_{\R^3} 
E(2l)^{-1}|\widehat{R}(l+k/2,l-k/2)|^2\,dl\right)^{1/2}\times\\
\times\left(\int_{\R^3} E(2l)|M(l+k/2,l-k/2)|^2\,dl\right)^{1/2}
\end{multline*} 
and finally
$$\int\frac{E(k)^2}{|k|^2}|\widehat{\rho_{1,0}}(k)|^2\,dk \leq 
\frac{1}{2^5\pi^3}\normR{R}^2\int_{|l|\leq\Lambda} 
\frac{1}{E(2l)E(l)^2}\,dl\leq 
\frac{\log\Lambda}{2^4\pi^2}\normR{R}^2$$
by $(\ref{estim_Log})$ which implies
$$\normr{\rho_{1,0}}\leq \frac{C_R(\log\Lambda)^{1/2}}{4\pi}\normQ{Q}.\qed$$

\subsubsection{Second order terms}
To simplify the presentation, we introduce the following notation:
\begin{equation}
S_{p,q}:=(4\pi)(2\pi)^{-3/p}\left(\int_{\R^3}\frac{du}{E(u)^q}\right)^{\frac{1}{p}},\quad 
S_p:=S_{p,p},\quad 
K_p:=\frac{1}{2\pi}\int_{-\ii}^{+\ii}\frac{d\eta}{E(\eta)^p}.
\label{S}
\end{equation} 
Let us recall the following inequality \cite[Theorem 4.1]{Simon}
\begin{equation}
\norm{f(x)g(-i\nabla)}_{\S_p}\leq 
(2\pi)^{-3/p}\norm{f}_{L^p(\R^3)}\norm{g}_{L^p(\R^3)},
\label{Simon}
\end{equation}
which implies
$$\norm{\frac{1}{|D_0|^{a}}f}_{\S_{p}} \leq \frac{S_{p,a 
p}}{4\pi}\norm{f}_{L^p(\R^3)}.$$

On the other hand, we shall often use the following trick
\begin{eqnarray*}
(E(p)^2+\eta^2)(E(q)^2+\eta^2) & = & 
(E(p)E(q))^2+(E(p)^2+E(q)^2)\eta^2+\eta^4\\
  & \geq & \frac{1}{4}E(p+q)^2+\frac{1}{2}E(p+q)^2\eta^2\\
  & \geq & \frac{1}{4}E(p+q)^2E(\eta)^2
\end{eqnarray*}
by Lemma \ref{peetre}. This implies
\begin{equation}
\label{trick}
\frac{1}{\sqrt{E(p)^2+\eta^2}\sqrt{E(q)^2+\eta^2}}\leq \frac{2}{E(p+q)E(\eta)}.
\end{equation}

\bigskip

Recall now that we have $Q_2=Q_{2,0}+Q_{1,1}+Q_{0,2}$ with
\begin{eqnarray*}
Q_{2,0} 	& = & -\frac{1}{2\pi} 
\int_{-\ii}^{+\ii}d\eta\frac{1}{D^0+i\eta}R\frac{1}{D^0+i\eta}R\frac{1}{D^0+i\eta}\\
Q_{1,1} & = & \frac{1}{2\pi} 
\int_{-\ii}^{+\ii}d\eta\frac{1}{D^0+i\eta}R\frac{1}{D^0+i\eta}\phi'\frac{1}{D^0+i\eta}\\
 & & \qquad+\frac{1}{2\pi}\int_{-\ii}^{+\ii}d\eta\frac{1}{D^0+i\eta}\phi'\frac{1}{D^0+i\eta}R\frac{1}{D^0+i\eta}\\
Q_{0,2} & = & -\frac{1}{2\pi} 
\int_{-\ii}^{+\ii}d\eta\frac{1}{D^0+i\eta}\phi'\frac{1}{D^0+i\eta}\phi'\frac{1}{D^0+i\eta}.
\end{eqnarray*}
We shall now treat each term separately.

\begin{lemma}
\label{order_2}
We have $\rho_{0,2}=0$ and the following estimates:
$$\normQ{Q_{2,0}}\leq 2^{5/2}K_{3/2}(C_R)^2\normQ{Q}^2,\qquad 
\normQ{Q_{1,1}}\leq 4S_6C_6K_{3/2}C_R \normQ{Q}\normr{\rho'},$$
$$\normQ{Q_{0,2}}\leq 2\sqrt{10}S_{6,5}C_M C_{6}\normr{\rho'}^2,$$
$$ \normr{\rho_{2,0}}\leq \frac{S_6C_6(C_R)^2}{\pi}\normQ{Q}^2,\qquad 
\normr{\rho_{1,1}}\leq 
4\frac{S_{6,5}C_MC_6C_R}{\pi}\normr{\rho'}\normQ{Q},$$
and so
$$\norm{(Q_2,\rho_2)}\leq \kappa_2 \norm{(Q,\rho')}^2,$$
with
$$\kappa_2=C_{Q_2}C_R\sqrt{2}+2\sqrt{\pi}C_{\rho_2},$$
$$C_{Q_2}=\max\left( 2^{3/2}K_{3/2}\, ,\, 
\frac{S_6C_6K_{3/2}}{\sqrt{2\pi}}\, , \, 
\frac{\sqrt{5}S_{6,5}C_MC_6}{\pi\sqrt{2}}\right)$$
$$C_{\rho_2}=\max\left( \frac{S_6C_6}{2\pi}\, , \, 
\frac{S_{6,5}C_MC_6}{\pi^{3/2}\sqrt{2}}\right),$$
where $C_M$ is a constant defined in Lemma \ref{estim_MQ}.
\end{lemma}

\noindent{\it Proof.} {\bf Step 1 : Estimates on the exchange term $Q_2$.}

\noindent $\bullet\ $ $Q_{2,0}$. To estimate $Q_{2,0}$, we write
$$|\widehat{Q_{2,0}}(p,q)|  \leq \frac{1}{2\pi} 
\int_{-\ii}^{+\ii}d\eta\int_{\R^3}\,dp_1\frac{|\widehat{R}(p,p_1)|}{\sqrt{E(p)^2+\eta^2}}
\frac{|\widehat{R}(p_1,q)|}{\sqrt{E(p_1)^2+\eta^2}}\frac{1}{\sqrt{E(q)^2+\eta^2}}$$
and so by $(\ref{trick})$
$$|\widehat{Q_{2,0}}(p,q)| \leq \frac{2^{3/2}}{2\pi} 
\int_{-\ii}^{+\ii}\frac{d\eta}{E(\eta)^{3/2}E(p+q)^{1/2}}\int_{\R^3}\frac{|\widehat{R}(p,p_1)|}{E(p+p_1)^{1/2}}
\frac{|\widehat{R}(p_1,q)|}{E(p_1+q)^{1/2}}\,dp_1$$
which implies
\begin{multline*}
E(p-q)E(p+q)^{1/2}|\widehat{Q_{2,0}}(p,q)| \leq 
2^{5/2}K_{3/2}\int_{\R^3}\frac{E(p-p_1)|\widehat{R}(p,p_1)|}{E(p+p_1)^{1/2}}\times\\
\times\frac{E(p_1-q)|\widehat{R}(p_1,q)|}{E(p_1+q)^{1/2}}\,dp_1
\end{multline*} 
and finally
$$\normQ{Q_{2,0}}\leq 2^{5/2}K_{3/2}\normR{R}^2\leq 
2^{5/2}K_{3/2}(C_R)^2\normQ{Q}^2.$$

\bigskip

\noindent $\bullet\ $ $Q_{1,1}$. We treat for instance
$$Q_{1,1}' :=  \frac{1}{(2\pi)^{5/2}} 
\int_{-\ii}^{+\ii}d\eta\frac{1}{D^0+i\eta}R\frac{1}{D^0+i\eta}\phi'\frac{1}{D^0+i\eta}$$
and use the same method to obtain
\begin{multline*}
E(p-q)E(p+q)^{1/2}|\widehat{Q'_{1,1}}(p,q)| \leq 
\frac{4}{(2\pi)^{5/2}} \int_{-\ii}^{+\ii}\frac{d\eta}{E(\eta)}\times\\
\times\int_{\R^3}\frac{E(p-p_1)}{E(p+p_1)^{1/2}}|\widehat{R}(p,p_1)|
\frac{E(p_1-q)}{(E(p_1)^2+\eta^2)^{1/4}(E(q)^2+\eta^2)^{1/4}}|\widehat{\phi'}(p_1-q)|\,dp_1.
\end{multline*}
This means that
$$\normQ{Q_{1,1}'}\leq \frac{4}{2\pi} 
\int_{-\ii}^{+\ii}\frac{d\eta}{E(\eta)}\norm{R'\frac{1}{(|D_0|^2+\eta^2)^{1/4}}f\frac{1}{(|D_0|^2+\eta^2)^{1/4}}}_{\S_2}$$
where we have introduced $R'$ and $f$ defined by
$$\widehat{R'}(p,q):=\frac{E(p-q)}{E(p+q)^{1/2}}|\widehat{R}(p,q)|,\ 
\ \ \widehat{f}(k):=E(k)|\widehat{\phi'}(k)|.$$
But now
\begin{eqnarray*}
\norm{R'\frac{1}{(|D_0|^2+\eta^2)^{\frac{1}{4}}}f\frac{1}{(|D_0|^2+\eta^2)^{\frac{1}{4}}}}_{\S_2}
& \leq & 
\norm{R'}_{\S_2}\norm{\frac{1}{(|D_0|^2+\eta^2)^{\frac{1}{4}}}f\frac{1}{(|D_0|^2+\eta^2)^{\frac{1}{4}}}}_{\S_\ii}\\
  & \leq & 
\normR{R}\norm{\frac{1}{(|D_0|^2+\eta^2)^{\frac{1}{4}}}f\frac{1}{(|D_0|^2+\eta^2)^{\frac{1}{4}}}}_{\S_6}.
\end{eqnarray*}
If we now use inequality $(\ref{Simon})$, we obtain
\begin{eqnarray*}
\norm{\frac{1}{(|D_0|^2+\eta^2)^{\frac{1}{4}}}f\frac{1}{(|D_0|^2+\eta^2)^{\frac{1}{4}}}}_{\S_6} 
& \leq & 
\norm{\frac{1}{(|D_0|^2+\eta^2)^{\frac{1}{4}}}|f|^{1/2}}_{\S_{12}}^2\\
  & \leq & 
(2\pi)^{-1/2}\left(\int_{\R^3}\frac{du}{1+|u|^2+\eta^2)^{3}}\right)^{1/6}\norm{f}_{L^6}\\
  & = & \frac{S_6}{4\pi E(\eta)^{1/2}}\norm{f}_{L^6}.
\end{eqnarray*}
Finally since
\begin{equation}
\label{estim_6}
\norm{f}_{L^6}\leq C_6\norm{\nabla f}_{L^2}=C_6\normp{\phi'},
\end{equation}
and $\normp{\phi'}=(4\pi)\normr{\rho'}$, we obtain
$$\normQ{Q_{1,1}'}\leq \frac{4S_6C_6}{2\pi} 
\int_{-\ii}^{+\ii}\frac{d\eta}{E(\eta)^{3/2}}\normR{R}\normr{\rho'}$$
and
$$\normQ{Q_{1,1}}\leq 8S_6C_6K_{3/2}C_R \normQ{Q}\normr{\rho'}.$$

\bigskip

\noindent $\bullet\ $ $Q_{0,2}$. Unfortunately, the method used above 
cannot be applied to $Q_{0,2}$. In this case, we have to calculate 
this term explicitely. We can write
$$Q_{0,2} 
=\sum_{\epsilon_1,\epsilon_2,\epsilon_3\in\{\pm\}}Q_{0,2}^{\epsilon_1\epsilon_2\epsilon_3}$$
where for instance (by a residuum formula)
$$\widehat{Q_{0,2}^{+++}}(p,q)=\widehat{Q_{0,2}^{---}}(p,q)=0,$$
$$\widehat{Q_{0,2}^{+--}}(p,q) =(2\pi)^{-3}
\int_{\R^3}dp_1\frac{\Lambda^+(p)\widehat{\phi'}(p-p_1)\Lambda^-(p_1)\widehat{\phi'}(p_1-q)\Lambda^-(q)}{(E(p)+E(q))(E(p)+E(p_1))},$$
$$\widehat{Q_{0,2}^{+-+}}(p,q) =(2\pi)^{-3}
\int_{\R^3}dp_1\frac{\Lambda^+(p)\widehat{\phi'}(p-p_1)\Lambda^-(p_1)\widehat{\phi'}(p_1-q)\Lambda^+(q)}{(E(p)+E(p_1))(E(q)+E(p_1))},$$ 
and similar formulas for the other 
$Q_{0,2}^{\epsilon_1\epsilon_2\epsilon_3}$. We now treat for instance 
$\widehat{Q_{0,2}^{+--}}$.
Using $(\ref{eq:estim_sum})$, we may obtain
\begin{multline*}
E(p-q)E(p+q)^{1/2}|\widehat{Q_{0,2}^{+--}}| \leq 2(2\pi)^{-3} 
\int_{\R^3}\,dp_1\frac{E(p-p_1)}{E(p+p_1)^{2/3}}\times\\
\times\left|\Lambda^+(p)\widehat{\phi'}(p-p_1)\Lambda^-(p_1)\right|\times
\frac{E(p_1-q)|\widehat{\phi'}(p_1-q)|}{E(p_1)^{1/3}E(q)^{1/2}}dp_1.
\end{multline*}
So, we may write
$$\normQ{Q_{0,2}^{+--}}\leq 
2\norm{M_{f}\frac{1}{|D_0|^{1/3}}f\frac{1}{|D_0|^{1/2}}}_{\S_2}\leq 
2\norm{M_{f}}_{\S_2}\norm{\frac{1}{|D_0|^{1/3}}f\frac{1}{|D_0|^{1/2}}}_{\S_\ii},$$
where
\begin{equation}
\label{MQ}
\widehat{M_f}(p,q):=(2\pi)^{-3/2}\frac{\widehat{f}(p-q)\left|\Lambda^+(p)\Lambda^-(q)\right|}{E(p+q)^{2/3}}.
\end{equation}
\begin{lemma}
\label{estim_MQ}
When $M_f$ is defined by formula $(\ref{MQ})$, then
$$\norm{M_f}_{\S_2}\leq \frac{C_M}{4\pi} 
\left(\int_{\R^3}|k|^2|\widehat{f}(k)|^2dk\right)^{1/2}$$
where 
$C_M:=2\left(\int_0^\ii\frac{t^2\,dt}{E(2t)^{4/3}E(t)^2}\right)^{1/2}\simeq 
2.1589$.
\end{lemma}

\noindent {\it Proof.} We have 
$\left|\Lambda^+(p)\Lambda^-(q)\right|^2=\Tr{\Lambda^+(p)\Lambda^-(q)}$ 
and so, by $(\ref{eq:estim_M})$,
\begin{eqnarray*}
\int\!\!\!\int\,dp\,dq|\widehat{M_f}(p,q)|^2 & \leq & 
(2\pi)^{-3}\int\!\!\!\int dp\,dq 
\frac{|p-q|^2|\widehat{f}(p-q)|^2}{2E((p+q)/2)^2E(p+q)^{4/3}}\\
  & \leq & (2\pi)^{-3}\int 
dk|k|^2|\widehat{f}(k)|^2\int\frac{du}{2E(2u)^{4/3}E(u)^2}\\
  & \leq & (2\pi)^{-2} \left(\int 
dk|k|^2|\widehat{f}(k)|^2\right)\int_0^\ii\frac{t^2\,dt}{E(2t)^{4/3}E(t)^2}.\qed
\end{eqnarray*}

Finally, since by $(\ref{estim_6})$
$$\norm{\frac{1}{|D_0|^{1/3}}f\frac{1}{|D_0|^{1/2}}}_{\S_\ii}\leq\norm{\frac{1}{|D_0|^{1/3}}f\frac{1}{|D_0|^{1/2}}}_{\S_6}\leq\frac{S_{6,5}}{4\pi}\norm{f}_{L^6},$$
we obtain
$$\normQ{Q_{0,2}^{+--}}  \leq  2S_{6,5}C_M C_{6}\normr{\rho'}^2.$$
This result is immediately extended to the others terms and since we 
can prove 
$$|Q|^2=|Q^{++-}+Q^{+--}|^2+|Q^{-++}+Q^{--+}|^2+|Q^{+-+}|^2+|Q^{-+-}|^2,$$
we arrive at
$$\normQ{Q_{0,2}}\leq 2\sqrt{10}S_{6,5}C_M C_{6}\normr{\rho'}^2.$$

\bigskip

{\bf Step 2 : Estimates on the density $\rho_2$.} Let us now treat 
the density $\rho_2$. The general idea of the proof is to estimate 
$\pscal{\rho_2,\zeta}$ in terms of the norm $\normd{\zeta}$ by using
$$|\pscal{\rho,\zeta}|=\left|\mathrm{Tr}\left(Q\zeta\right)\right|  = 
\left|\mathrm{Tr}\left(\widehat{Q\zeta}\right)\right|
  =  \left|\int_{\R^3}\Tr{\widehat{Q\zeta}}(p,p)\,dp\right|\leq 
\int_{\R^3}|\widehat{Q\zeta}(p,p)|dp.$$
This can be done if we know that $Q\zeta\in\S_1$. But we have
\begin{multline*}
\norm{Q\zeta}_{\S_1} = \norm{Q|D_0|^2\frac{1}{|D_0|^2}\zeta}_{\S_1} 
\leq  \norm{Q|D_0|^2}_{\S_2}\norm{\frac{1}{|D_0|^2}\zeta}_{\S_2}\\ \leq 
E(\Lambda)^2\norm{Q}_{\S_2} \frac{S_{2,4}}{4\pi}\norm{\zeta}_{L^2},
\end{multline*}  
showing that $Q\zeta\in\S_1$ when $\zeta\in L^2$. So, in what 
follows, we shall assume that $\zeta\in\EuFrak{C}'\cap L^2$ and 
prove a bound depending only on $\normd{\zeta}$. By the density of 
$\EuFrak{C}'\cap L^2$ in $\EuFrak{C}'$, this will give us a bound 
on $\normr{\rho}$.\\

Let us remark first that $\rho_{0,2}$ vanishes. Indeed we have
$$\widehat{\rho_{0,2}}(k)=\frac{1}{(2\pi)^{3/2}}\int_{|p|\leq\Lambda} 
\Tr{\widehat{Q_{0,2}}(p+k/2,p-k/2)}dp$$
and
\begin{multline*}
\Tr{\widehat{Q_{0,2}}(p,q)}= \frac{1}{(2\pi)^{4}} 
\int_{-\ii}^{+\ii}d\eta\int_{\R^3}\,dp_1\mathrm{Tr}_{\C^4}\left(\frac{1}{D_0(p)+i\eta}\widehat{\phi'}(p-p_1)\times\right.\\
\left.\times\frac{1}{D_0(p_1)+i\eta}\widehat{\phi'}(p_1-q)\frac{1}{D_0(q)+i\eta}\right)\\
= \frac{1}{(2\pi)^{4}} 
\int_{-\ii}^{+\ii}d\eta\int_{\R^3}\,dp_1\frac{\widehat{\phi'}(p-p_1)\widehat{\phi'}(p_1-q)}{\sqrt{E(p)^2+\eta^2}\sqrt{E(p_1)^2+\eta^2}\sqrt{E(q)^2+\eta^2}}\times\\
\times\mathrm{Tr}_{\C^4}\bigg[(D_0(p)-i\eta))(D_0(p_1)-i\eta)(D_0(q)-i\eta)\bigg].
\end{multline*}
Now the terms linear in the Dirac matrices are traceless and the 
remaining terms are odd in $\eta$ and vanish after integration. This 
can be easily generalized to $\rho_{0,2k}$ for all $k$, and is known 
as \emph{Furry's Theorem} in the physics literature \cite{Furry_thm}.

\bigskip

\noindent $\bullet\ $ $\rho_{2,0}$. We use here a method similar to 
what we have done above. We estimate for some 
$\zeta\in\EuFrak{C}'\cap L^2$ and $Q_\zeta:=Q_{2,0}\zeta$
\begin{eqnarray*}
|\widehat{Q_\zeta}(p,p)| & \leq & 
(2\pi)^{-5/2}\int_{-\ii}^{+\ii}d\eta\int\!\!\!\int 
\frac{|\widehat{R}(p,p_1)|\,|\widehat{R}(p_1,p_2)|\,|\widehat{\zeta}(p_2-p)|\,dp_1\,dp_2}{\sqrt{E(p)^2+\eta^2}\sqrt{E(p_1)^2+\eta^2}\sqrt{E(p_2)^2+\eta^2}},\\
  & \leq & 
4(2\pi)^{-5/2}\int_{-\ii}^{+\ii}\frac{d\eta}{E(\eta)}\int\!\!\!\int 
dp_1\,dp_2
\frac{E(p-p_1)|\widehat{R}(p,p_1)|\,E(p_1-p_2)|\widehat{R}(p_1,p_2)|}{E(p+p_1)^{1/2}E(p_1+p_2)^{1/2}}\times\\
  & & 
\qquad\times\frac{|\widehat{\zeta}(p_2-p)|}{E(p_2-p)(E(p)^2+\eta^2)^{1/4}(E(p_2)^2+\eta^2)^{1/4}},\\
& \leq & 
4(2\pi)^{-5/2}\int_{-\ii}^{+\ii}\frac{d\eta}{E(\eta)}\int\!\!\!\int 
dp_1\,dp_2 
\frac{R'(p,p_1)R'(p_1,p_2)\widehat{\zeta'}(p_2-p)}{(E(p)^2+\eta^2)^{1/4}(E(p_2)^2+\eta^2)^{1/4}}
\end{eqnarray*}
where 
$\widehat{R'}(p,q)=\frac{E(p-q)|\widehat{R}(p,q)|}{E(p+q)^{1/2}}$ and 
$\widehat{\zeta'}(k)=E(k)^{-1}\widehat{\zeta}(k)$. This means that
\begin{eqnarray*}
|\pscal{\rho_{2,0},\zeta}| & \leq & 
\frac{4}{2\pi}\int_{-\ii}^{+\ii}\frac{d\eta}{E(\eta)}\norm{R'R'\frac{1}{(|D_0|^2+\eta^2)^{1/4}}\zeta'\frac{1}{(|D_0|^2+\eta^2)^{1/4}}}_{\S_1}\\
  & \leq & 
\frac{4S_6}{(2\pi)(4\pi)}\int_{-\ii}^{+\ii}\frac{d\eta}{E(\eta)^{3/2}}\norm{R'}_{\S_2}^2\norm{\zeta'}_{L^6}\leq 
\frac{S_6C_6(C_R)^2K_{3/2}}{\pi}\normQ{Q}^2\normd{\zeta}
\end{eqnarray*}
by $(\ref{estim_6})$, showing that
$$\normr{\rho_{2,0}}\leq \frac{S_6C_6(C_R)^2}{\pi}K_{3/2}\normQ{Q}^2.$$

\bigskip

\noindent $\bullet\ $ $\rho_{1,1}$. Unfortunately, as for $Q_{0,2}$, 
we have to calculate $\rho_{1,1}$ explicitely.
Let us start for instance with  $\rho_{1,1}^{+--}$, the density 
associated with one of the two terms of $Q_{1,1}$
$$(2\pi)^{-3/2}\int_{\R^3}dp_1\frac{\Lambda^+(p)\widehat{R}(p-p_1)\Lambda^-(p_1)\widehat{\phi'}(p_1-q)\Lambda^-(q)}{(E(p)+E(q))(E(p)+E(p_1))}.$$
We use the same method  as above and estimate for some 
$\zeta\in\EuFrak{C}'\cap L^2$ the term
$$\widehat{Q_\zeta}(p,p)=(2\pi)^{-3}\int\!\!\!\int 
dp_1\,dp_2\frac{\Lambda^+(p)\widehat{R}(p,p_1)\Lambda^-(p_1)\widehat{\phi'}(p_1-p_2)\Lambda^-(p_2)}{(E(p)+E(p_2))(E(p)+E(p_1))}\widehat{\zeta}(p_2-p),$$
by
\begin{eqnarray*}
|\widehat{Q_\zeta}(p,p)| & \leq & (2\pi)^{-3}\int\!\!\!\int 
\frac{|\Lambda^+(p)\widehat{R}(p,p_1)\Lambda^-(p_1)|}{E(p+p_1)^{1/2}}\frac{|\widehat{\phi'}(p_1-p_2)|}{E(p_1)^{1/2}E(p_2)^{1/3}}\times\\
  & & \qquad\qquad 
\times\frac{|\widehat{\zeta}(p_2-p)|\times|\Lambda^-(p_2)\Lambda^+(p)|}{E(p+p_2)^{2/3}}dp_1\,dp_2,\\
  & \leq & 2(2\pi)^{-3/2}\int\!\!\!\int 
\frac{E(p-p_1)|\Lambda^+(p)\widehat{R}(p,p_1)\Lambda^-(p_1)|}{E(p+p_1)^{1/2}}\times\\
  & & \qquad\times\frac{\widehat{f}(p_1-p_2)}{E(p_1)^{1/2}E(p_2)^{1/3}}
\widehat{M_{\zeta'}}(p_2,p)dp_1\,dp_2,
\end{eqnarray*}
with $\widehat{f}(k):=E(k)|\widehat{\phi'}(k)|$ and 
$\widehat{\zeta'}(k):=|\widehat{\zeta}(k)|/E(k)$. Now
$$|\widehat{Q_\zeta}(p,p)|  \leq  2(2\pi)^{-3/2}\int\!\!\!\int \widehat{R_1}(p,p_1)\frac{\widehat{f}(p_1-p_2)}{E(p_1)^{1/2}E(p_2)^{1/3}}\widehat{M_{\zeta'}}(p_2,p)\,dp_1\,dp_2$$
with
$$\widehat{R_1}(p,p_1):=\frac{E(p-p_1)|\Lambda^+(p)\widehat{R}(p,p_1)\Lambda^-(p_1)|}{E(p+p_1)^{1/2}}.$$
We thus have
\begin{eqnarray*}
|\pscal{\rho_{1,1}^{+--},\zeta}| & \leq & 2\norm{R_1 
\frac{1}{|D_0|^{1/2}}f\frac{1}{|D_0|^{1/3}} M_{\zeta'}}_{\S^1}\\
  & \leq  & 
2\normR{\Lambda^+R\Lambda^-}\norm{M_{\zeta'}}_{\S_2}\norm{\frac{1}{|D_0|^{1/2}}f\frac{1}{|D_0|^{1/3}}}_{\S_\ii}\\
  & \leq  & 
2\normR{\Lambda^+R\Lambda^-}\norm{M_{\zeta'}}_{\S_2}\norm{\frac{1}{|D_0|^{1/2}}f\frac{1}{|D_0|^{1/3}}}_{\S_6}\\
  & \leq  & 
\frac{S_{6,5}C_MC_6}{2\pi}\normR{\Lambda^+R\Lambda^-}\normd{\zeta}\normr{\rho'},
\end{eqnarray*}
and finally
$$\normr{\rho_{1,1}^{+--}}\leq 
\frac{S_{6,5}C_MC_6}{2\pi}\normR{\Lambda^+R\Lambda^-}\normr{\rho'}.$$

We now treat $\rho_{1,1}^{+-+}$ and estimate
$$\widehat{Q_\zeta}(p,p)=(2\pi)^{-3}\int\!\!\!\int 
dp_1\,dp_2\frac{\Lambda^+(p)\widehat{R}(p,p_1)\Lambda^-(p_1)\widehat{\phi'}(p_1-p_2)\Lambda^+(p_2)}{(E(p)+E(p_1))(E(p_1)+E(p_2))}\widehat{\zeta}(p_2-p),$$
by
\begin{multline*}
|\widehat{Q_\zeta}(p,p)|  \leq  2(2\pi)^{-3}\int\!\!\!\int 
\frac{E(p-p_1)|\Lambda^+(p)\widehat{R}(p,p_1)\Lambda^-(p_1)|}{E(p+p_1)^{1/2}}\times\\
\times\frac{\widehat{f}(p_1-p_2)|\Lambda^-(p_1)\Lambda^+(p_2)|}{E(p_1+p_2)^{2/3}}
\frac{\widehat{\zeta'}(p_2-p)}{E(p_2)^{5/6}}dp_1\,dp_2.
\end{multline*}
Using the same argument as above, we arrive at
$$\normr{\rho_{1,1}^{+-+}}\leq 
\frac{S_{6,5}C_MC_6}{2\pi}\normR{\Lambda^+R\Lambda^-}\normr{\rho'}.$$
To treat $\rho_{1,1}^{++-}$, we remark that
\begin{multline*}
\frac{1}{(E(p)+E(p_2))(E(p_1)+E(p_2))}\leq 
\frac{1}{(E(p)+E(p_1))(E(p_1)+E(p_2))}\\
+\frac{1}{(E(p)+E(p_1))(E(p)+E(p_2))}
\end{multline*}
and use the same estimates as above to get
$$\normr{\rho_{1,1}^{++-}}\leq 
\frac{S_{6,5}C_MC_6}{\pi}\normR{\Lambda^+R\Lambda^-}\normr{\rho'}.$$
Finally, since 
$\sum_{\epsilon_1,\epsilon_2\in\{\pm\}}\normR{\Lambda^{\epsilon_1}R\Lambda^{\epsilon_2}}^2=\normR{R}^2$, 
we end up with
$$\normr{\rho_{1,1}} \leq 
2\frac{S_{6,5}C_MC_6}{\pi}\normr{\rho'}\left(\sum_{\epsilon_1,\epsilon_2\in\{\pm\}}\normR{\Lambda^{\epsilon_1}R\Lambda^{\epsilon_2}}\right) 
\leq 4\frac{S_{6,5}C_MC_6C_R}{\pi}\normr{\rho'}\normQ{Q}.$$

\subsubsection{The general $n^{th}$ order case}
Now that we have explained how the proof works for the second order, 
let us estimate the general $n^{th}$ order term.

\begin{lemma}
\label{n_order_case}
We have the following estimates
$$\forall n\geq 3,\qquad \normQ{Q_n}\leq nK_{\frac{n}{2}} C_Q 
\norm{(Q,\rho')}^n,\ \mbox{ with }\ 
C_Q=\sqrt{2}\left(\frac{S_6C_6}{2\sqrt{\pi}}\right)^3,$$
$$\forall n\geq 5,\qquad \normr{\rho_{n}}\leq 
nK_{\frac{n+1}{2}}C_\rho\norm{(Q,\rho')}^n,\ \mbox{ with }\ 
C_\rho=\frac{S_6C_6}{4\pi}\left(\frac{S_6C_6}{2\sqrt{\pi}}\right)^5,$$
$$\normr{\rho_{4}}\leq C_{\rho_4}\norm{(Q,\rho')}^4,\ \mbox{ with }\ 
C_{\rho_4}:=\frac{K_{2}S_6C_6}{\pi}\left(\frac{S_6C_6}{2\sqrt{\pi}}\right)^2.$$
Therefore,
$$\forall n\geq 4,\qquad \norm{(Q_n,\rho_n)}\leq \kappa_n \norm{(Q,\rho')}^n$$
with
$$\kappa_4= 4C_RK_{2}C_Q\sqrt{2}+2\sqrt{\pi}C_{\rho_4},\qquad 
\kappa_n= 
nC_RK_{\frac{n}{2}}C_Q\sqrt{2}+2nK_{\frac{n+1}{2}}C_{\rho}\sqrt{\pi}.$$
\end{lemma}

Remark that it can be proved that 
$K_n\sim_{n\to\ii}\frac{C}{\sqrt{n}}$, which gives the claimed 
behaviour for $\kappa_n$ as $n\to\ii$.

\noindent{\it Proof.} {\bf Step 1 : Estimates on the exchange term $Q_n$.}

\noindent $\bullet\ $ {\it $Q_{k,l}$ with $k\geq 1$ and $k+l=n\geq 3$}.
Recall that
$$Q_{k,l}=\frac{(-1)^{l+1}}{2\pi} \sum_{I\cup J =\{1,...,n\},\ 
|I|=k,\ |J|=l} 
\int_{-\ii}^{+\ii}d\eta\frac{1}{D^0+i\eta}\prod_{j=1}^n 
\left(R_j\frac{1}{D^0+i\eta}\right),$$
where $R_j=R$ if $j\in I$ and $R_j=\phi'$ if $j\in J$. For the sake 
of simplicity, we treat only
$$Q'_{k,l} = \frac{1}{2\pi} 
\int_{-\ii}^{+\ii}d\eta\frac{1}{D^0+i\eta} 
\left(R\frac{1}{D^0+i\eta}\right)^k\left(\phi'\frac{1}{D^0+i\eta}\right)^l.$$
We have
\begin{multline*}
|\widehat{Q'_{k,l}}(p,q)| \leq  \frac{1}{(2\pi)^{1+\frac{3l}{2}}} 
\int_{-\ii}^{+\ii}d\eta\int\!\!\cdots\!\!\int 
\frac{1}{(E(p)^2+\eta^2)^{1/2}} |\widehat{R}(p,p_1)|\times\\
\times\frac{1}{(E(p_1)^2+\eta^2)^{1/4}}\prod_{j=1}^{k-1}\left(\frac{1}{(E(p_j)^2+\eta^2)^{1/4}} 
|\widehat{R}(p_j,p_{j+1})|\frac{1}{(E(p_{j+1})^2+\eta^2)^{1/4}}\right)\times\\
\times\prod_{j=k}^{n-2}\left(\frac{1}{(E(p_j)^2+\eta^2)^{1/4}} 
|\widehat{\phi'}(p_j-p_{j+1})|\frac{1}{(E(p_{j+1})^2+\eta^2)^{1/4}}\right)\times\\
\times\frac{1}{(E(p_{n-1})^2+\eta^2)^{1/4}} 
|\widehat{\phi'}(p_{n-1}-q)|\frac{1}{(E(q)^2+\eta^2)^{1/2}} 
\,dp_1\cdots dp_{n-1}
\end{multline*}
so by $(\ref{trick})$,
\begin{multline*}
E(p-q)E(p+q)^{1/2}|\widehat{Q'_{k,l}}(p,q)| \leq 
\frac{2^\frac{k+1}{2}E(p-q)}{(2\pi)^{1+\frac{3l}{2}}} 
\int_{-\ii}^{+\ii}\frac{d\eta}{E(\eta)^{\frac{k+1}{2}}}\int\!\!\cdots\!\!\int\frac{|\widehat{R}(p,p_1)|}{E(p+p_1)^{1/2}}\times\\
\times \prod_{j=1}^{k-1}\frac{|\widehat{R}(p_j,p_{j+1})|}{E(p_j+p_{j+1})^{1/2}}
\prod_{j=k}^{n-2}\left(\frac{1}{(E(p_j)^2+\eta^2)^{1/4}} 
|\widehat{\phi'}(p_j-p_{j+1})|\frac{1}{(E(p_{j+1})^2+\eta^2)^{1/4}}\right)\times\\
\times\frac{1}{(E(p_{n-1})^2+\eta^2)^{1/4}} 
|\widehat{\phi'}(p_{n-1}-q)|\frac{1}{(E(q)^2+\eta^2)^{1/4}} 
\,dp_1\cdots dp_{n-1}.
\end{multline*}
Now if we use the easy generalization of $(\ref{eq:peetre_sum})$,
$$E(p-q)\leq E(p-p_1)+E(p_1-p_2)+\cdots + E(p_{n-2}-p_{n-1})+E(p_{n-1}-q),$$
we obtain by a similar argument as before
\begin{multline*}
\normQ{Q'_{k,l}}\leq \frac{2^\frac{k+1}{2}}{2\pi}\left[k(C_R)^k(4\pi 
C_\ii)^l\left(\int_{-\ii}^\ii\frac{d\eta}{E(\eta)^{l+\frac{k+1}{2}}}\right)\right.\\
\left.+ l(C_R)^k(4\pi 
C_\ii)^{l-1}S_6C_6\left(\int_{-\ii}^\ii\frac{d\eta}{E(\eta)^{l+\frac{k}{2}}}\right)\right]\normQ{Q}^k\normr{\rho'}^l.
\end{multline*}
To obtain this result, we have estimated each term containing a 
$\phi'$ by using
$$\norm{\frac{1}{(|D_0|^2+E(\eta)^2)^{1/4}}\phi'\frac{1}{(|D_0|^2+E(\eta)^2)^{1/4}}}_{\S_\ii} 
\leq \frac{1}{E(\eta)}\norm{\phi'}_{L^\ii}\leq 
\frac{C_\ii}{E(\eta)}\normp{\phi'}$$ and when $E(p_j-p_{j+1})$ 
appears in front of a $\widehat{\phi'}(p_j-p_{j+1})$ (i.e. when 
$j\geq k$), by using
\begin{multline*}
\norm{\frac{1}{(|D_0|^2+E(\eta)^2)^{1/4}}f\frac{1}{(|D_0|^2+E(\eta)^2)^{1/4}}}_{\S_\ii} 
\\
\leq 
\norm{\frac{1}{(|D_0|^2+E(\eta)^2)^{1/4}}f\frac{1}{(|D_0|^2+E(\eta)^2)^{1/4}}}_{\S_6} 
\leq
\frac{S_6}{E(\eta)^{1/2}4\pi}\norm{f}_{L^6}.
\end{multline*}
So we have
\begin{eqnarray*}
\normQ{Q'_{k,l}} & \leq & 2^\frac{k+1}{2}\left[k(C_R)^k(4\pi 
C_\ii)^lK_{l+\frac{k+1}{2}}+ l(C_R)^k(4\pi 
C_\ii)^{l-1}S_6C_6K_{l+\frac{k}{2}}\right]\normQ{Q}^k\normr{\rho'}^l\\
  & \leq & 2^\frac{k+1}{2}n(C_R)^k(4\pi 
C_\ii)^lK_{l+\frac{k}{2}}\max\left(1, \frac{S_6C_6}{4\pi 
C_\ii}\right)\normQ{Q}^k\normr{\rho'}^l
\end{eqnarray*}
which implies
$$\normQ{Q_{k,l}} \leq \left( _k^n\right)2^\frac{k+1}{2}n(C_R)^k(4\pi 
C_\ii)^lK_{l+\frac{k}{2}}\max\left(1, \frac{S_6C_6}{4\pi 
C_\ii}\right)\normQ{Q}^k\normr{\rho'}^l.$$

\bigskip

\noindent $\bullet\ $ {\it $Q_{0,n}$ with $n\geq 3$}.
Recall that
$$Q_{0,n} =\frac{(-1)^{n+1}}{2\pi} 
\int_{-\ii}^{+\ii}d\eta\frac{1}{D^0+i\eta}\left(\phi'\frac{1}{D^0+i\eta}\right)^n$$
so that
\begin{multline*}
E(p-q)E(p+q)^{1/2}|\widehat{Q_{0,n}}(p,q)| \leq 
\frac{E(p-q)}{(2\pi)^{1+\frac{3n}{2}}} 
\int_{-\ii}^{+\ii}\frac{d\eta}{E(\eta)^{\frac{1}{2}}}\int\!\!\cdots\!\!\int 
\,dp_1\cdots dp_{n-1}\times\\
\times\frac{|\widehat{\phi'}(p-p_1)|}{(E(p)^2+\eta^2)^{1/4}(E(p_1)^2+\eta^2)^{1/4}}\times\\
\times\prod_{j=1}^{n-2}\left(\frac{1}{(E(p_j)^2+\eta^2)^{1/4}} 
|\widehat{\phi'}(p_j-p_{j+1})|\frac{1}{(E(p_{j+1})^2+\eta^2)^{1/4}}\right)\times\\
\times\frac{1}{(E(p_{n-1})^2+\eta^2)^{1/4}} 
|\widehat{\phi'}(p_{n-1}-q)|\frac{1}{(E(q)^2+\eta^2)^{1/4}}.
\end{multline*}
We now use $(\ref{estim_6})$ to bound for some $f_1$, $f_2$ and $f_3$
\begin{equation}
\norm{\prod_{j=1}^3\left(\frac{1}{(|D_0|^2+\eta^2)^{1/4}}f_j\frac{1}{(|D_0|^2+\eta^2)^{1/4}}\right)}_{\S^2}\leq 
\frac{(S_6)^3}{E(\eta)^{3/2}(4\pi)^3}\prod_{j=1}^3\norm{f_j}_{L^6}
\label{3S6}
\end{equation}
to obtain
$$\normQ{Q_{0,n}}\leq 
n\frac{\sqrt{2}}{2\pi}\left(\int_{-\ii}^\ii\frac{d\eta}{E(\eta)^{2+(n-3)}}\right)(S_6C_6)^3(4\pi 
C_\ii)^{n-3}\normr{\rho'}^n$$
or
$$\normQ{Q_{0,n}}\leq nK_{n-1}\sqrt{2}(S_6C_6)^3(4\pi 
C_\ii)^{n-3}\normr{\rho'}^n.$$

Finally, we can write for instance (recall that $C_\ii=1/(2\sqrt{\pi})$)
\begin{equation}
\normQ{Q_n}\leq nK_{n/2} C_Q 
\left(C_R\sqrt{2}\normQ{Q}+2\sqrt{\pi}\normr{\rho'}\right)^n
\end{equation}
where
$$C_Q=\sqrt{2}\max\left(1, \frac{S_6C_6}{4\pi C_\ii} , 
\left(\frac{S_6C_6}{4\pi 
C_\ii}\right)^3\right)=\sqrt{2}\frac{(S_6C_6)^3}{8\pi^{3/2}},$$
and since $K_{n/2}\geq K_{n-1}$ when $n\geq 2$.

\bigskip

\noindent {\bf Step 2 : Estimates on the density $\rho_n$.}\\
$\bullet\ $ {\it $\rho_{k,l}$ with $k\geq 2$ and  $n\geq 3$}. As 
before we treat for instance the density $\rho_{k,l}'$ of the 
$Q'_{k,l}$ where the $k$ $R$'s are on the left and the $l$ $\phi'$'s 
are on the right. For some fixed $\zeta\in\EuFrak{C}'\cap L^2$, we 
introduce $Q_\zeta:=Q'_{k,l}\zeta$. We thus estimate
\begin{multline*}
|\widehat{Q_\zeta}(p,p)| \leq  \frac{1}{(2\pi)^{1+\frac{3(l+1)}{2}}} 
\int_{-\ii}^{+\ii}d\eta\int\!\!\cdots\!\!\int 
\frac{1}{(E(p)^2+\eta^2)^{1/4}} |\widehat{R}(p,p_1)|\times\\
\times\frac{1}{(E(p_1)^2+\eta^2)^{1/4}}\prod_{j=1}^{k-1}\left(\frac{1}{(E(p_j)^2+\eta^2)^{1/4}} 
|\widehat{R}(p_j,p_{j+1})|\frac{1}{(E(p_{j+1})^2+\eta^2)^{1/4}}\right)\times\\
\times\prod_{j=k}^{n-1}\left(\frac{1}{(E(p_j)^2+\eta^2)^{1/4}} 
|\widehat{\phi'}(p_j-p_{j+1})|\frac{1}{(E(p_{j+1})^2+\eta^2)^{1/4}}\right)\times\\
\times\frac{1}{(E(p_{n})^2+\eta^2)^{1/4}}|\widehat{\zeta}(p_n-p)|\frac{1}{(E(p)^2+\eta^2)^{1/4}} 
\,dp_1\cdots dp_{n}.
\end{multline*}
We now use as before
$$|\widehat{\zeta}(p_n-p)|\leq 
\frac{|\widehat{\zeta}(p_n-p)|}{E(p_n-p)}\big(E(p-p_1)+E(p_1-p_2)+\cdots+E(p_{n-1}-p_n)\big)$$
to obtain
$$|\pscal{\rho_{k,l}',\zeta}|\leq 
n\frac{2^{k/2}S_6C_6}{4\pi}(C_R)^k(4\pi 
C_\ii)^l\left(\frac{1}{2\pi}\int_{-\ii}^\ii\frac{d\eta}{E(\eta)^{l+\frac{k+1}{2}}}\right)\normQ{Q}^k\normr{\rho'}^l\normd{\zeta}$$
and so
$$\normr{\rho_{k,l}}\leq 
n\left(^n_k\right)\frac{S_6C_6}{4\pi}(C_R\sqrt{2})^k(4\pi 
C_\ii)^lK_{l+\frac{k+1}{2}}\normQ{Q}^k\normr{\rho'}^l.$$

\bigskip

\noindent $\bullet\ $ {\it $\rho_{1,l}$ with $l\geq 2$}. We may treat 
for instance with the same notation as before
\begin{multline*}
|\widehat{Q_\zeta}(p,p)| \leq  \frac{1}{(2\pi)^{1+\frac{3n}{2}}} 
\int_{-\ii}^{+\ii}d\eta\int\!\!\cdots\!\!\int 
\frac{1}{(E(p)^2+\eta^2)^{1/4}} |\widehat{R}(p,p_1)|\times\\
\times\frac{1}{(E(p_1)^2+\eta^2)^{1/4}}\prod_{j=1}^{n-1}\left( 
\frac{1}{(E(p_j)^2+\eta^2)^{1/4}} 
|\widehat{\phi'}(p_j-p_{j+1})|\frac{1}{(E(p_{j+1})^2+\eta^2)^{1/4}}\right)\times\\
\times\frac{1}{(E(p_{n})^2+\eta^2)^{1/4}}|\widehat{\zeta}(p_n-p)|\frac{1}{(E(p)^2+\eta^2)^{1/4}} 
\,dp_1\cdots dp_{n}.
\end{multline*}
We now use $(\ref{3S6})$ and obtain
$$|\pscal{\rho_{1,l}',\zeta}|\leq 
n\frac{2^{1/2}(S_6C_6)^3}{4\pi}C_R(4\pi 
C_\ii)^{l-2}\left(\frac{1}{2\pi}\int_{-\ii}^\ii\frac{d\eta}{E(\eta)^{l-2+\frac{4}{2}}}\right)\normQ{Q}\normr{\rho'}^l\normd{\zeta}$$
and so
$$\normr{\rho_{1,l}}\leq 
n\left(^n_{n-1}\right)\frac{2^{1/2}(S_6C_6)^3K_{l}}{4\pi}C_R(4\pi 
C_\ii)^{l-2}\normQ{Q}\normr{\rho'}^l.$$

\bigskip

\noindent $\bullet\ $ {\it $\rho_{0,l}$ with $l\geq 5$}. We want to estimate
\begin{multline*}
|\widehat{Q_\zeta}(p,p)| \leq  \frac{1}{(2\pi)^{1+\frac{3(n+1)}{2}}} 
\int_{-\ii}^{+\ii}d\eta\int\!\!\cdots\!\!\int 
\frac{1}{(E(p)^2+\eta^2)^{1/4}} |\widehat{\phi'}(p-p_1)|\times\\
\times\frac{1}{(E(p_1)^2+\eta^2)^{1/4}}\prod_{j=1}^{n-1}\left(\frac{1}{(E(p_j)^2+\eta^2)^{1/4}} 
|\widehat{\phi'}(p_j-p_{j+1})|\frac{1}{(E(p_{j+1})^2+\eta^2)^{1/4}}\right)\times\\
\times\frac{1}{(E(p_{n})^2+\eta^2)^{1/4}}|\widehat{\zeta}(p_n-p)| 
\frac{1}{(E(p)^2+\eta^2)^{1/4}}\,dp_1\cdots dp_{n}.
\end{multline*}
Since there are at least $6$ functions, we may use $(\ref{3S6})$ 
twice and obtain
$$|\pscal{\rho_{0,l}',\zeta}|\leq n\frac{(S_6C_6)^6}{4\pi}(4\pi 
C_\ii)^{l-5}\left(\frac{1}{2\pi}\int_{-\ii}^\ii\frac{d\eta}{E(\eta)^{l-2}}\right)\normr{\rho'}^n\normd{\zeta}$$
and so
$$\normr{\rho_{0,l}}\leq n\frac{(S_6C_6)^6}{4\pi}(4\pi 
C_\ii)^{l-5}K_{l-2}\normr{\rho'}^n.$$

Now since $K_{n-2}\leq K_{(n+1)/2}$ when $n\geq 5$, we obtain
\begin{equation}
\normr{\rho_{n}}\leq 
nK_{\frac{n+1}{2}}C_\rho\left(C_R\sqrt{2}\normQ{Q}+2\sqrt{\pi}\normr{\rho'}\right)^n
\end{equation}
with
$$C_\rho:=\frac{S_6C_6}{4\pi}\max\left(1,\left(\frac{C_6S_6}{2\sqrt{\pi}}\right)^2,\left(\frac{C_6S_6}{2\sqrt{\pi}}\right)^5\right)=\frac{S_6C_6}{4\pi}\left(\frac{C_6S_6}{2\sqrt{\pi}}\right)^5.$$

For $\rho_4$, we notice that $\rho_{0,4}=0$ for the same reason as 
$\rho_{0,2}$, and that $K_{2}\geq K_{5/2}$. Therefore we obtain
\begin{equation}
\normr{\rho_{4}}\leq 
C_{\rho_4}\left(C_R\sqrt{2}\normQ{Q}+2\sqrt{\pi}\normr{\rho'}\right)^4
\end{equation}
with
$$C_{\rho_4}:=\frac{4K_{2}S_6C_6}{4\pi}\max\left(1,\left(\frac{C_6S_6}{2\sqrt{\pi}}\right)^2\right)=\frac{S_6C_6K_2}{\pi}\left(\frac{C_6S_6}{2\sqrt{\pi}}\right)^2.\qed$$

\subsubsection{The third order density $\rho_3$}
\begin{lemma}
\label{order_3}
We have
$$\normr{\rho_3}\leq C_{\rho_3} 
\left(C_R\sqrt{2}\normQ{Q}+2\sqrt{\pi}\normr{\rho'}\right)^3$$
and therefore
$$\norm{(Q_3,\rho_3)}\leq \kappa_3\norm{(Q,\rho')}^3$$
with
$$\kappa_3=3C_RK_{3/2}C_Q\sqrt{2}+2\sqrt{\pi}C_{\rho_3},\qquad 
C_{\rho_3}=\frac{15C_MS_6(S_{6,4})^2(C_6)^4}{\pi(4\pi C_\ii)^3}.$$
\end{lemma}

\noindent {\it Proof.} Notice that thanks to the previous proof, 
we already have some estimates on $\rho_{3,0}$, $\rho_{2,1}$ and 
$\rho_{1,2}$. It remains to study $\rho_{0,3}$.
As before and as in \cite{HS}, we have to compute $\rho_{0,3}$ 
explicitely by a residuum formula. We thus write
$$\rho_{0,3}=\sum_{\epsilon_1,...,\epsilon_4\in\{\pm\}}\rho_{0,3}^{\epsilon_1\epsilon_2\epsilon_3\epsilon_4}$$
with an obvious definition.

\bigskip

\noindent $\bullet$ Let us treat first $\rho_{0,3}^{+---}$. We thus 
fix some $\zeta\in\EuFrak{C}'\cap L^2$ and estimate the term
\begin{multline*}
\widehat{Q_\zeta}(p,p)=(2\pi)^{-6}\int\!\!\!\int\!\!\!\int 
dp_1\,dp_2\,dp_3\frac{\Lambda^+(p)\widehat{\phi'}(p-p_1)\Lambda^-(p_1)}{E(p)+E(p_1)}\times\\
\times\frac{\widehat{\phi'}(p_1-p_2)\Lambda^-(p_2)\widehat{\phi'}(p_2-p_3)\Lambda^-(p_3)}{(E(p)+E(p_2))(E(p)+E(p_3))}\widehat{\zeta}(p_3-p),
\end{multline*}
by
\begin{multline*}
|\widehat{Q_\zeta}(p,p)|  \leq  (2\pi)^{-6}\int\!\!\!\int\!\!\!\int 
dp_1\,dp_2\,dp_3\frac{\left|\Lambda^+(p)\widehat{\phi'}(p-p_1)\Lambda^-(p_1)\right|}{E(p+p_1)^{2/3}}\times\\
\times\frac{\left|\widehat{\phi'}(p_1-p_2)\right| 
\left|\widehat{\phi'}(p_2-p_3)\right|}{E(p_1)^{1/3}E(p_2)E(p)}\left|\widehat{\zeta}(p_3-p)\right|.
\end{multline*}
So if we follow the method used above, we obtain
$$\normr{\rho_{0,3}^{+---}}\leq \frac{3 
C_MS_6(S_{6,4})^2(C_6)^4}{4\pi}\normr{\rho'}^3.$$

Now, it is easily seen that $\rho_{0,3}^{-+++}$, $\rho_{0,3}^{---+}$, 
$\rho_{0,3}^{+++-}$, $\rho_{0,3}^{-+--}$, $\rho_{0,3}^{+-++}$, 
$\rho_{0,3}^{--+-}$ and $\rho_{0,3}^{++-+}$ can be treated by exactly 
the same method.

\bigskip

\noindent $\bullet$ Let us now treat for instance 
$\rho_{0,3}^{++--}$. Thanks to the residuum formula, we have to study
\begin{multline*}
\widehat{Q_\zeta}(p,p)=(2\pi)^{-6}\int\!\!\!\int\!\!\!\int 
dp_1\,dp_2\,dp_3\Lambda^+(p)\widehat{\phi'}(p-p_1)\Lambda^+(p_1)\widehat{\phi'}(p_1-p_2)\Lambda^-(p_2)\\
\widehat{\phi'}(p_2-p_3)\Lambda^-(p_3)\widehat{\zeta}(p_3-p)
\times\left(\frac{1}{(E(p)+E(p_2))(E(p_1)+E(p_2))(E(p_1)+E(p_3))}+\right.\\
\left.\frac{1}{(E(p)+E(p_2))(E(p)+E(p_3))(E(p_1)+E(p_3))}\right).
\end{multline*}
If we now use the same method as above for each of the two terms of 
this sum, we arrive at
$$\normr{\rho_{0,3}^{++--}}\leq 2\frac{3 
C_MS_6(S_{6,4})^2(C_6)^4}{4\pi}\normr{\rho'}^3.$$
This is easily generalized to the study of $\rho_{0,3}^{--++}$, 
$\rho_{0,3}^{+-+-}$, $\rho_{0,3}^{-+-+}$, $\rho_{0,3}^{+--+}$ and 
$\rho_{0,3}^{-++-}$.

Summing now all these terms, we obtain
$$\normr{\rho_{0,3}}\leq 20\frac{3 
C_MS_6(S_{6,4})^2(C_6)^4}{4\pi}\normr{\rho'}^3= \frac{15 
C_MS_6(S_{6,4})^2(C_6)^4}{\pi}\normr{\rho'}^3$$
and
\begin{equation}
\normr{\rho_3}\leq C_{\rho_3} 
\left(C_R\sqrt{2}\normQ{Q}+2\sqrt{\pi}\normr{\rho'}\right)^3,
\end{equation}
with
\begin{eqnarray*}
C_{\rho_3}  & = & 3\frac{S_6C_6}{4\pi}\max\left(K_{3+1/2}\, ,\, K_3\, 
,\,  K_2\left(\frac{S_6C_6}{4\pi C_\ii}\right)^2 , 
\frac{20C_M(S_{6,4})^2(C_6)^3}{(4\pi C_\ii)^3} \right)\\
 & = & \frac{15C_MS_6(S_{6,4})^2(C_6)^4}{\pi(4\pi C_\ii)^3}.
\end{eqnarray*} 
\qed

\subsubsection{List of constants}\label{def_constantes}
We have used many constants in this proof. A summary which should help the reader to follow our arguments is provided in Table 1.
\begin{table}\label{table_constantes}
\begin{center}
\begin{tabular}{|l|l|}
\hline
Constant & defined in\\
\hline
$C_6$, $C_\ii$ & Lemma \ref{injections}\\
$C_R$ & Equation \eqref{CR} in Lemma \ref{estim_RQ}\\
$\kappa_1(\Lambda)$ & Equation \eqref{Kappa_1} in Proposition \ref{estimates}\\
$S_{p,q}$, $S_p$, $K_p$ & Equation \eqref{S}\\
$\kappa_2$, $C_{Q_2}$, $C_{\rho_2}$ & Lemma \ref{order_2}\\
$C_M$ & Lemma \ref{estim_MQ}\\
$C_Q$, $C_\rho$, $C_{\rho_4}$ & Lemma \ref{n_order_case}\\
$\kappa_n$, $n\geq4$ & Lemma \ref{n_order_case}\\
$C_{\rho_3}$, $\kappa_3$ & Lemma \ref{order_3}\\
\hline
\end{tabular} 
\caption{Constants used in the proof of Theorem \ref{existence}}
\end{center} 
\end{table}

\section*{Appendix: Derivation of the BDF energy}
\label{derivation}
In this section, we recall some basics about the second-quantization 
in no-photon QED and explain how the BDF energy 
$\mathcal{E}$ is derived from this theory, as a 
mean-field approximation. We mainly follow the method of 
Chaix-Iracane \cite{Chaix, Chaix_th}, but with the notation of 
\cite{Thaller, HF1, HeSi}. See also \cite{HS} for more details 
concerning the polarization of the vacuum.
To simplify the presentation, we introduce $P^0_-:=P^0$ and $P^0_+:=1-P^0$.

\subsection*{Free particles, Fock space, free vacuum}
We first introduce $\mathcal{F}^{(1)}_+:=P^0_+\mathcal{H}_\Lambda$ 
and $\mathcal{F}^{(1)}_-:=CP^0_-\mathcal{H}_\Lambda$ which are called 
respectively the free electron and the free positron state subspace. 
$C$ is the charge-conjugation operator defined by 
$C\psi=i\beta\alpha_2\overline{\psi}$. We define 
$\mathcal{F}^{(0)}_+=\mathcal{F}^{(0)}_-=\C$ and 
$\mathcal{F}^{(n)}_+=\bigwedge_{k=1}^n\mathcal{F}^{(1)}_+$, 
$\mathcal{F}^{(m)}_-=\bigwedge_{k=1}^m\mathcal{F}^{(1)}_-$ for 
$n,m\geq 1$. The space of $n$ free electrons and $m$ free positrons 
is then defined by $\mathcal{F}^{(n,m)}=\mathcal{F}^{(n)}_+\otimes 
\mathcal{F}^{(m)}_-$ and the associated Fock space is
\begin{equation}
\label{fockspace}
\mathcal{F}:=\bigoplus_{n,m=0}^{\ii}\mathcal{F}^{(n,m)}.
\end{equation}

For any $f\in\mathcal{H}_\Lambda$, the free electron (resp. positron) 
annihilation and creation operators $a_0(f)$ and $a_0^\ast(f)$ (resp. 
$b_0(f)$ and $b_0^*(f)$) are defined as usually \cite{Thaller}.
They fulfill the Canonical Anti-commutation Relations
\begin{equation}
\label{acr}\{a_0(f),a_0(g)\}= \{a_0(f),b_0(g)\} =\{b_0(f),b_0(g)\}=0,
\end{equation}
\begin{equation}
\label{acr2}
\{a_0(f),a_0^*(g)\}=\pscal{f,P^0_+ g}, \qquad 
\{b_0^*(f),b_0(g)\}=\pscal{f,P^0_- g},
\end{equation}
where $\pscal{\cdot,\cdot}$ denotes the usual scalar product of 
$L^2(\R^3,\C^4)$.

The \emph{free vacuum state} $\Omega_0$, a unit vector spanning 
$\mathcal{F}^{(0,0)}=\C$, is uniquely characterized up to a phase 
factor by the properties
\begin{equation}
\label{free_vac}
\norm{\Omega_0}_{\cal F}=1,\quad a_0(f)\Omega_0=0\quad \text{and}\quad b_0(f)\Omega_0=0,
\end{equation}
for all $f\in\mathcal{H}_\Lambda$.

\medskip

The \emph{field operator} $\Psi(f)$ is defined on the Fock space 
$\mathcal{F}$ by
$$\Psi(f)=a_0(f)+b_0^\ast(f).$$
In terms of $\Psi(f)$, the CAR become, for all $(f,g)\in\mathcal{H}_\Lambda^2$,
$$\{\Psi(f),\Psi(g)\}= \{\Psi^*(f),\Psi^*(g)\}=0,\qquad 
\{\Psi(f),\Psi^\ast(g)\}=\pscal{f,g} 1,$$

\subsection*{Dressed particles and vacuum}
In this description, the free electrons and positrons are defined 
with respect to the projector $P^0$, or equivalently the splitting 
$\mathcal{H}_\Lambda=\mathcal{H}_-^0\oplus\mathcal{H}_+^0$. We want 
now to change this definition and introduce the \emph{dressed 
electrons and positrons}. To this end, we fix a new projector $P$ on 
$\mathcal{H}_\Lambda$, use again the notation $P_-:=P$ and 
$P_+:=1-P$, and introduce the dressed particle annihilation operators
\begin{equation}
\label{rel_Psi2}
a_P(f):=\Psi(P_+f),\qquad b_P(f):=\Psi^\ast(P_-f).
\end{equation}
Similar formula can be given for the dressed particle creation 
operators $a_P^\ast$ and $b_P^\ast$.
These dressed operators satisfy the same CAR as for the free 
operators $(\ref{acr})$ $(\ref{acr2})$. We also introduce the dressed 
electrons and positrons state subspaces
$$\mathcal{H}_+^P=(1-P)\mathcal{H}_\Lambda,\qquad 
\mathcal{H}_-^P=P\mathcal{H}_\Lambda.$$

Now, the main question is to know if there exists a \emph{dressed 
vacuum} $\Omega_P$ in the Fock space $\mathcal{F}$. This state has to 
be a solution to the analogue of $(\ref{free_vac})$
\begin{equation}
\label{dres_vac}
\norm{\Omega_P}_{\cal F}=1,\quad a_P(f)\Omega_P=0\quad \text{ and  }\quad b_P(f)\Omega_P=0
\end{equation}
for all $f\in\mathcal{H}_\Lambda$. The answer is given by the celebrated
\begin{thm}[Shale-Stinespring Theorem \cite{ShSt}]\label{shale} There 
exists a dressed vacuum $\Omega_P$ in the Fock space $\mathcal{F}$ 
satisfying $(\ref{dres_vac})$ if and only if $P-P^0$ is a 
Hilbert-Schmidt operator. In this case, $\Omega_P$ is unique up to a 
phase factor.
\end{thm}
There are many proofs of this Theorem in the literature, see, e.g., 
\cite{Thaller,Scharf1,Rui} and the references in \cite{FS}.
This result explains why we assumed in the previous section that 
$P-P^0\in\S_2(\mathcal{H}_\Lambda)$. Notice that $\Omega_P$ can be 
expressed as a rotation of the bare vacuum in the Fock space, 
$\Omega_P=\mathbb{U}\Omega_0$, $\mathbb{U}$ being called a Bogoliubov 
transformation. An explicit formula for $\Omega_P$ can be found in a 
lot of papers \cite{Thaller,Scharf1,Rui,SS,Seipp,FS}.

The charge of the dressed vacuum $\Omega_P$ can be easily 
computed\footnote{Notice that the charge of the dressed vacuum can also be 
easily obtained by using the explicit formula of $\Omega_P$, which 
immediately shows that it is an integer \cite{Thaller,KS2,Seipp,SS}.}
$$\pscal{\Omega_P,\mathcal{Q}\,\Omega_P}=\tr[P^0_+(P-P^0)P^0_+]+\tr[P^0_-(P-P^0)P^0_-]=\str_{P^0}(P-P^0),$$
justifying our study of Section \ref{supertrace}. Here $\cal Q$ is 
the charge operator defined on $\mathcal{F}$ by \cite[Formula 
(10.52)]{Thaller}
$$\mathcal{Q}=\sum_{i\in\Z\setminus\{0\}}\bigg\{a_0^*(f_i)a_0(f_i)-b_0^*(f_i)b_0(f_i)\bigg\}=\sum_{i\geq1}a_0^*(f_i)a_0(f_i)-\sum_{i\leq-1}b_0^*(f_i)b_0(f_i),$$
where $(f_i)_{i\geq1}$ is an orthonormal basis of $\mathcal{H}^0_+$ 
and $(f_i)_{i\leq-1}$ is an orthonormal basis of $\mathcal{H}^0_-$.

\subsection*{Second-quantized Hamiltonian}
In the physics literature, the creation and annihilation operators 
are defined differently. For instance, instead of $a_0^\ast(f)$ which 
creates an electron in the state $P^0_+f$, the operator $a_0^\ast(x)$ 
which creates an electron at $x$ is formally used, where 
$a_0^\ast(x)=\sum_{i=1}^\ii a_0^\ast(f_i)f_i(x)$, $(f_i)_{i\geq 1}$ 
being an orthonormal basis of $\mathcal{H}_+^0$. The operators 
$a_P(x)$ and $b_P(x)$ are defined similarly.
We shall now use this formalism which is also the one of \cite{HS}. 
Formally, the CAR $(\ref{acr},\ref{acr2})$ are equivalent to
\begin{equation}
\label{facr1}
\{a_0(x),a_0(y)\}= \{a_0(x),b_0(y)\}=\{a_0^*(x),b_0(y)\}=\{b_0(x),b_0(y)\}=0,
\end{equation}
\begin{equation}
   \label{facr2}
   \{a_0(x),a_0^*(y)\}=P^0_+ (x,y), \qquad \{b_0^*(x),b_0(y)\}=P^0(x,y).
\end{equation}

We now start with writing down the formal unregularized no-photon Hamiltonian
\begin{equation}
   \label{ie}
   \Hh_\mathrm{ur} = \int dx\, \Psi^*(x)
   D^{\alpha\phi}\Psi(x)
   + \frac{\alpha}2 \int dx \int dy
   \frac{\Psi^*(x)\Psi(x)\Psi^*(y)\Psi(y)}{|\bx - \by|},
\end{equation}
which acts on the Fock space $\mathcal{F}$. As explained for instance 
in \cite{Thaller}, the free vacuum may not belong to the domain 
of this formally defined operator. Therefore, the expression \eqref{ie} is renormalized by using a procedure which 
is called ``normal ordering", denoted by double dots $:-:_{P^0}$. In 
each product of annihilation and creation operators, the $a_0^\ast$ 
and $b_0^\ast$ are moved to the left as if they anticommute with the 
$a_0$ and $b_0$. For instance
\begin{eqnarray}
:\Psi^*(x)\Psi(y):_{P^0} & = & a^*_0(x) a_0(y) + a_0^*(y) b_0(x) + 
b_0(x)a_0(y) - b^*_0(y)b_0(x) \nonumber \\
 & = &  \Psi^*(x) \Psi(y) - P^0(x,y).\label{normo}
\end{eqnarray}

As a first step we thus regularize $\Hh_\mathrm{ur}$ as done in 
Chaix and Iracane \cite[Sections 3.5 and 4.1]{Chaix}, namely we 
normal order with respect
to the free projector $P^0$,
\begin{equation}
   \label{rh}
   \Hh = \int dx\,: \Psi^*(x)
   D^{\alpha\phi}\Psi(x) :_{P^0}
   + \frac{\alpha}2 \int dx \int dy
   \frac{:\Psi^*(x)\Psi(x)\Psi^*(y)\Psi(y):_{P^0}}{|\bx - \by|}.
\end{equation}
This kind of regularization, which follows ideas of Dirac \cite{D1,D}, is standard in QED. It corresponds to the subtraction of the energy of the free Dirac 
sea, and the interaction energy with the free Dirac sea. A physical justification is given in \cite[Section 3]{HS}, on the basis of
two guiding principles formulated by Weisskopf in \cite{Weis}. The same choice is 
made in other studies dealing with vacuum polarization, for instance 
\cite{Chaix,Chaix_th,Scharf1,KS2,SS,Seipp,FS,ABHTS}.
However, a better choice might be possible (see the paper \cite{LS} by Lieb and Siedentop, who propose another
translation-invariant reference for normal ordering, in the absence of external field).\medskip

Now we can express $\Hh$ in terms of $:-:_P$ for some other $P$. 
Using \eqref{normo} we obtain the reordering relations
\begin{equation}
:\Psi^*(x)\Psi(y):_{P^0} =  :\Psi^*(x)\Psi(y):_P + Q(x,y)
\end{equation}
and
\begin{multline*}
:\Psi^*(x)\Psi(x) \Psi^*(y)\Psi(y) :_{P^0}\ =\ :\Psi^*(x)\Psi(x) 
\Psi^*(y)\Psi(y) :_P \\ +2 :\Psi^*(x)\Psi(x):_P \Tr{Q(y,y)}
  - 2:\Psi^*(x)\Psi(y):_P  Q(x,y)\\ + \Tr{Q(x,x)}\Tr{Q(y,y)} - |Q(x,y)|^2
\end{multline*}
where $Q = P - P^0$.  Therefore we can rewrite $\Hh$ with respect to an arbitrary dressed 
vacuum $P$ \cite[formula $(4.3)$]{Chaix},
\begin{multline}
\label{H}\mathbb{H}= \int :\Psi^\ast(x)D^{\alpha\phi}\Psi(x):_Pdx 
+\frac{\alpha}{2} \int\!\!\!\int 
\frac{:\Psi^\ast(x)\Psi(x)\Psi^\ast(y)\Psi(y):_P}{|x-y|}dx\,dy\\
+\alpha\int\!\!\!\int 
\frac{:\Psi^\ast(x)\Psi(x):_P\Tr{Q(y,y)}}{|x-y|}dx\,dy 
-\alpha\int\!\!\!\int 
\frac{:\Psi^\ast(x)\Psi(y):_PQ(x,y)}{|x-y|}dx\,dy\\
+\str_{P^0}(D^\phi Q)+\frac{\alpha}{2}\int\!\!\!\int 
\frac{\Tr{Q(x,x)}\Tr{Q(y,y)}}{|x-y|}dx\,dy 
-\frac{\alpha}{2}\int\!\!\!\int \frac{|Q(x,y)|^2}{|x-y|}dx\,dy.
\end{multline}
The last line represents the energy of the dressed vacuum $P$ 
measured with respect to $P^0$, whereas in the second line the vacuum 
polarization potentials appear.

\subsection*{Restriction to Bogoliubov-Dirac-Fock states}
We now follow \cite{Chaix} and restrict ourselves to 
Bogoliubov-Dirac-Fock type states. In this approximation method, a 
dressed vacuum $P$ is first chosen such that 
$P-P^0\in\S_2(\mathcal{H})$. Then a $BDF$ state is simply a Slater 
determinant made with $n$ electrons and $m$ positrons defined with 
respect to the dressed vacuum $P$ ($n,m\geq 0$ are not fixed in this 
theory). This is a state of $\mathcal{F}$ which takes the form
$$\psi=a_P^*(f_1)\cdots a_P^*(f_n)\, b_P^*(g_1)\cdots b_P^*(g_m)\, \Omega_P,$$
where $(f_1,...,f_n)\in(\mathcal{H}_+^P)^n$ and 
$(g_1,...,g_m)\in(\mathcal{H}_-^P)^m$ are such that 
$\pscal{f_i,f_j}=\delta_{ij}$, $\pscal{g_i,g_j}=\delta_{ij}$,
and $\Omega_P$ is the dressed vacuum in $\mathcal{F}$ obtained by 
Theorem \ref{shale}. Since it is easily seen that $\psi=\Omega_{P'}$ 
where
$$P'=P+\gamma,\qquad \gamma=\sum_{i=1}^n|f_i\rangle\langle 
f_i|-\sum_{j=1}^m|g_j\rangle\langle g_j|,$$
we obtain immediately from \eqref{H}
$$\pscal{\psi|\Hh|\psi} = \pscal{\Omega_{P'}|\Hh|\Omega_{P'}}= 
\mathcal{E}(P'-P^0)= \mathcal{E}(P-P^0+\gamma).$$
In \cite[formula $(4.8)$]{Chaix}, this formula is expanded like in \eqref{separation} in terms of the vacuum density matrix
$Q=P-P^0$ and the density $\gamma$ of the dressed particles.

\bibliographystyle{amsplain}

\end{document}